\documentclass[a4paper,english,twocolumn,nofootinbib,pre,superscriptaddress]{revtex4-2}
\usepackage{amsmath,amsfonts}
\usepackage{physics}
\usepackage{amssymb}
\usepackage{mathrsfs}
\usepackage{ulem}
\usepackage{graphicx}
\usepackage{xcolor}
\usepackage{braket}
\usepackage{hyperref}

\newcommand{\be}{\begin{equation}}
\newcommand{\ee}{\end{equation}}

\newcommand{\E}{\mathcal{E}}
\newcommand{\ov}{\overline}

\newcommand{\ra}{\rangle}
\newcommand{\la}{\langle}

\newcommand{\ww}{\widetilde}

\newcommand{\HH}{{\mathcal{H}}}

\newcommand{\LL}{{\mathcal{L}}}
\newcommand{\M}{{\mathcal{M}}}

\makeatletter

\begin{document}

\title{
 An ETH-ansatz-motivated environmental-branch approach to 
 open quantum systems 
}

\author{Wen-ge Wang}\email{wgwang@ustc.edu.cn}
\affiliation{ Department of Modern Physics, University of Science and Technology of China,
	Hefei 230026, China}
\affiliation{CAS Key Laboratory of Microscale Magnetic Resonance,
University of Science and Technology of China, Hefei 230026, China}
\affiliation{Anhui Center for fundamental sciences in theoretical physics, Hefei 230026, China}

\date{\today}

\begin{abstract}

In this paper, a method is developed for the study of  a generic small central quantum system,  which is locally coupled to an environment as a many-body quantum chaotic system that satisfies the eigenstate thermalization hypothesis (ETH) ansatz. The approach is based on properties of environmental branches of the total system's state, the overlaps of which give the reduced density matrix (RDM) of the central system. To study evolution of the RDM within a finite time period, the period is divided into a series of short intervals, within each of which the RDM is computed by making use of a formal solution to the time evolution of the environmental branches. The expressions thus obtained are simplified by the ETH ansatz and, further, by decay of phase correlations among the environmental branches, the latter of which also originates from chaotic dynamics of the environment. This gives a generic method of deriving master equation. And, as an application, a master equation is derived in a simplest nontrivial case, which predicts a decoherence rate in agreement with that predicted by the random-matrix theory. Furhermore, the Born approximation, which is employed in the ordinary approach to master equation, can be justified within the proposed framework; and a Markovian feature is shown for the RDM's evolution in an effective sense. 
\end{abstract}

\maketitle

\section{Introduction}

\subsection{Motivation}

 Open quantum system is one of the most important and widely studied topics in modern physics.
 Measurable properties of such systems, characterized by expectation values of their observables,
 are given by their reduced density matrices (RDM).
 Due to entanglement of an open system with its environment,
 its RDM is usually not sufficient for determining its time evolution.
 However, as is known, under an environment such as a thermal bath satisfying certain conditions,
 evolution of the RDM approximately follows a so-called master equation \cite{breuer2002}.
 Typically, derivation of a master equation, particularly, that of a Lindblad form,  is based on
 certain approximations such as Born approximation and Markov approximation.

 The ordinary analytical framework does not, in a quantitative way, give dynamic justification on applicability of the Born and Markov approximations. 
 From the dynamic perspective, even the concept of quantum thermal bath is hypothetical,
 since it neglects dynamical correlations that may be induced by system-environment interactions.
 Indeed, in recent years, growing interest has been seen in various non-Markovian effects,
 for example, back-action of the central system on environment
 and its application in controlling coherence
 (see, e.g., Refs.\cite{GongJB12,addis2014coherence,roszak2015decoherence,
 zhang2015role,cakmak2017,guarnieri2018steady}).

 In the past few decades, it has been found that quantum chaotic systems may possess many
 of the properties typically ascribed to quantum thermal baths
 \cite{RBC06,Rigol-AiP16,Deutch-RPP18},
 especially for systems that satisfy the so-called eigenstate thermalization hypothesis (ETH) ansatz
 \cite{Rigol-AiP16,Deutch-RPP18,Deutch91,srednicki1994ET,srednicki-JPA96,srednicki1999ETH}.
 For the purpose of studying dynamics of open quantum systems,
 it is reasonable to expect that taking an environment as a dynamical system
 that satisfies ETH — particularly,
 a many-body quantum chaotic system, rather than a hypothetical thermal bath —
 may  be more beneficial.
 Indeed, recent studies have shown that, in the derivation of master equation,
 the ETH ansatz is helpful for arguing validity of the Markov approximation
 and useful in the calculation of correlation functions \cite{DonvSMGM-PRB25}.

 In the study of dynamic effects of environment,
 the most direct method is to decompose the total system's state $|\Psi\ra$ as
 $|\Psi\ra = \sum_\alpha |\alpha\ra |\phi^\E_\alpha\ra$,
 where $|\alpha\ra$ are the energy eigenstates of the central system
 and $|\phi^\E_\alpha\ra$ are the corresponding environmental branches.
 In fact, inner products of the environmental branches yield elements of the RDM, namely,
 $\la \phi_\beta^\E(t)|\phi_\alpha^\E(t)\ra = \rho^S_{\alpha \beta}(t)$.
 One question of interest is whether and how chaoticity of an environment,
 particularly the ETH ansatz,
 could be useful in the study of dynamic properties of the environmental branches
 and hence of the RDM.

 The above discussed approach has been found
 useful in the study of nondissipative system-environment interactions,
 \footnote{Here, dissipative interaction means that the system-environment interaction Hamiltonian
 is not commutable with the system's Hamiltonian. }
 in which evolution of the environmental branches is in fact 
 governed by a Schr\"{o}dinger-type dynamics under
 certain effective Hamiltonian (see, e.g., Refs.\cite{GPSS04,pra08-PS}).
 In this case, offdiagonal RDM elements take a form of
 the so-called quantum Loschmidt echo (LE), which
 gives a measure to the sensitivity  of quantum motion under small perturbation \cite{peres84}.
 For quantum chaotic systems,
 main decaying behaviors of the LE have been studied
 \cite{jal2001,CT02,jac2001,cuc2002,prosen2002,STB03,GPSZ06,WCL04,WL05}
 and can be directly used in the study of pure dephasing effects,
 resorting to neither the Born nor the Markov approximation \cite{GPSS04}.
 However, under a generic dissipative interaction,
 mutual influences among environmental branches make their evolution significantly more complex.

\subsection{Purpose and basic strategy of this paper}

 In this paper, we study a generic small central quantum system,
 which is locally coupled to an environment as a generic many-body quantum
 chaotic system that satisfies the ETH ansatz.
 One main purpose is to show that a master equation can be derived for the RDM of
 the central system under approximations,
 which are based on the chaotic dynamics of the environment,
 particularly those stated in the ETH ansatz.
 For this, one needs to compute the time evolution of 
 the environmental branches and, then, study effects of correlations 
 among the environmental branches, which are induced by the
 system-environment interaction.

 To achieve the above discussed goal, we are to derive
 a formal expression for the evolution of the environmental branches.
 Based on this, the following steps are proposed 
 for deriving a master equation.
 That is, firstly, a finite time period of interest is divided into a series 
 of short time intervals, within each of which the above mentioned 
 formal expression is effectively truncated at certain $k_{\rm tru}$-th order of terms.
 Secondly, the expression of the RDM thus obtained is simplified by
 making use of the ETH ansatz, as well as of some other approximations
 which also originate from chaotic dynamics of the environment.
 And, thirdly, summing up all nonnegligible contributions left, a master equation 
 may be derived.

 More specifically, the paper is organized as follows.
 The setup is given in Sec.\ref{sect-setup}, in which systems to be 
 studied are specified and basic contents of the ETH ansatz are recalled.
 In Sec.\ref{sect-branch-ETH}, basic contents of 
 the environmental-branch approach to the RDM evolution are discussed, 
 as well as preliminary analysis based on the ETH ansatz.
 In Sec.\ref{sect-finite-t-evolve}, a formal expression for the time evolution 
 of environmental branches is derived and is used for deriving an approximate 
 expression for a finite time evolution of the RDM, 
 with the help of the ETH ansatz.

 Based on results obtained above,
 a basic framework for deriving a master equation 
 is proposed in Sec.\ref{sect-basic-framework}.
 As an application of the framework, in Sec.\ref{sect-ME-ktru=2}, 
 the simplest nontrivial case of truncation with $k_{\rm tru}=2$ is studied,
 which results in a master equation.
 Prediction of the master equation for the decoherence rate 
 of pure dephasing is shown to be consistent with what is known 
 by the random-matrix theory (RMT).
 Furthermore, in Sec.\ref{sect-refinement},
 one treatment in the basic framework is improved, 
 giving rise to a refined framework.

 In Sec.\ref{sect-Born-Mark-appr}, within the proposed framework, 
 the Born approximation is justified and a Markovian feature is argued
 for the RDM evolution.
 Finally, conclusions and discussions are given in Sec.\ref{sect-conclusion}.

\section{Setup}\label{sect-setup}


 The central system to be studied is
 a generic relatively small quantum system, denoted by $S$.
 It is locally coupled to a huge environment as a generic many-body  quantum chaotic system denoted by $\E$, which satisfies the ETH ansatz.
 The particle number of the environment $\E$ is denoted by $N$.
 The Hilbert spaces of $S$ and $\E$ are denoted by $\HH_S$ and $\HH_\E$, respectively,
 with dimensions $d_S$ and $d_\E$.
 The Hamiltonian of the total composite system $S + \E$ is written as
\begin{align}\label{}
 H = H^S + H^\E + H^I,
\end{align}
 where $H^S$ and $H^\E$ indicate the self-Hamiltonians of $S$ and $\E$, respectively,
 which are determined in the limit of weak system-environment coupling.
 The interaction is local  and has a product form,
\begin{align}\label{}
 H^I = H^{IS} \otimes H^{I\E},
\end{align}
 where $H^{IS}$ is an arbitrary observable of the system $S$
 and $H^{I\E}$ is an arbitrary local observable of $\E$.

 Eigenstates of $H$ are denoted by $|n\ra$, with energies $E_n$;
 and, those of $H^S$ and $H^\E$ are denoted by $|\alpha\ra$ and $|i\ra$, respectively,
 with eigenenergies $e^S_\alpha$ and $e^\E_i$,
\begin{subequations}
\begin{align}\label{}
 & H|n\ra = E_n|n\ra,
 \\ & H^S|\alpha\ra = e^S_\alpha|\alpha\ra,
 \\ & H^\E|i\rangle = e_i^{\E}|i\rangle.
\end{align}
\end{subequations}
 All the eigenenergies are ordered by increasing energy.
 On the eigenbases, the system and environment parts of the interaction Hamiltonian elements are written as
$ H^{IS}_{\alpha \beta} = \la\alpha|H^{IS}|\beta\ra$ and $  H^{I\E}_{ij} = \la i | H^{I\E}|j\ra$, respectively.
 The total eigenstates are expanded as follows,
\begin{align}\label{}
 |n\rangle = \sum_{\alpha i}C_{\alpha i}^n|\alpha i\rangle.
\end{align}


 For an arbitrary observable $O$ of the environment,
 its matrix elements on the energy basis $\{ |i\ra \}$, $O_{ij} = \la i |O|j\ra$,
 can always be written in the following form,
\begin{equation}\label{ETH-0}
 O_{ij} = O(e^\E_i) \delta_{ij} + R_{ij},
\end{equation}
 where $O(e)$ is some smooth function,
 $R_{ii}$ represent deviation of the diagonal elements $O_{ii}$ from $O(e)$,
 and $R_{ij}$ of $i \ne j$ are the offdiagonal elements of $O$.
 The ETH ansatz states that the matrix of $O_{ij}$ of certain type of observable $O$ possesses
 a specific structure, as discussed below.

 Although the scope within which the ETH ansatz is applicable is still not completely clear,
 it is usually expected to include at least local observables of many-body quantum chaotic systems
 (see, e.g., reviews in Refs.\cite{Rigol-AiP16,Deutch-RPP18}).
 For such an observable $O$,
 the ETH ansatz conjectures that the diagonal function $O(e)$ varies slowly with $e$
 and the quantities $R_{ij}$ show certain random feature.
 More exactly, the ETH ansatz is written as \cite{srednicki-JPA96,srednicki1999ETH}
\begin{equation}\label{ETH}
 O_{ij} = O(e^\E_i) \delta_{ij} + \frac{1}{\sqrt{\rho_{\rm dos}^\E(e^0)}} f(e^0,\omega) r_{ij},
\end{equation}
 where $O(e)$ is a slowly-varying function,
 $\rho_{\rm dos}^\E (e)$ is the density of states of the environment, and
 $f(e^0, \omega)$ is a smooth function, with
 $e^0 = (e^\E_i + e^\E_j)/2$ and $\omega = e^\E_i - e^\E_j$.
 Here, $r_{ij} = r^*_{ji}$ are independent random variables,
 with normal distribution for $i\ne j$ (zero mean and unit variance).
 Originally, the term $1/\sqrt{\rho_{\rm dos}^\E}$  on the right-hand side (rhs) of Eq.(\ref{ETH})
 was written as $e^{-S(e^0)/2}$, where $S(e)$ indicates the thermodynamic entropy
 \cite{srednicki-JPA96,srednicki1999ETH};
 a point is that both $S(e)$ and $\rho_{\rm dos}^\E$ are proportional to the particle number $N$.

 Concerning the diagonal fluctuation, the variance of $r_{ii}$ was found
 deviating from that of $r_{ij}$ with $i\ne j$ by some factor denoted by $\eta_r$.
 Generically, it is expected that $\eta_r = 2$ in systems with the time reversal symmetry
 \cite{Sred-H-98,Eckhardt95,Wilkinson,Eckhardt-Main95,Castella_1996,pra22-PB-ETH}
 and $\eta_r = 1$ otherwise \cite{Sred-H-98};
 while, in some specific models,
 the value of $\eta$ may lie between $1$ and $2$ according to numerical simulations
 \cite{Steinigeweg_2013,Rigol_2017,Dymarsky_Gemmer_2020}.

 Significant correlations have been found in the quantity $R_{ij}$ by
 both numerical studies and analytical analysis
 \cite{Jiaozi22ETHdevi,Vidmar20-prl,Dymarsky22BoundETH,Nussinov22,Jiaozi-PRX25,
 pre25-ETHsc,ctp25-obs-corr,WW-ETH-conf},
 though the random matrix theory predicts a purely random feature of  $R_{ij}$.
 For example, high-order moments of $r_{ij}$ were found possessing correlations \cite{Jiaozi22ETHdevi}, and 
 the offdiagonal function $f(e^0,\omega)$ should possess a bounded feature 
 \cite{Dymarsky22BoundETH,WW-ETH-conf}.
 Indeed, in model numerical simulations, a banded structure of the function $f(e^0,\omega)$
 is usually observed with respect to the variable $\omega$.
 More exactly, with $|\omega|$ increasing from zero, the function decays relatively slowly within
 some region (showing a platform or somehow power-law decay), whose width is to be indicated as $w_f$, 
 and then decays exponentially beyond the region.
 In discussions to be given below, effects from high-order moments of $r_{ij}$ are not to be considered,
 while, the banded structure of the offdiagonal function $f$ will considered.

\section{Environmental branch formulation of RDM}\label{sect-branch-ETH}

 In this section, we discuss the basic form of the environmental-branch approach to
 the RDM evolution (Sec.\ref{sect-branch-RDM}).
 We also give a preliminary analysis from the perspective of
 the ETH ansatz (Sec.\ref{HIE-under-ETH}).

\subsection{Environmental-branch expression of RDM}\label{sect-branch-RDM}

 To get an expression of the RDM in terms of environmental branches,
 let us consider an arbitrary initial state $|\Psi(t_0)\ra$, expanded as $ |\Psi(t_0)\ra = \sum_n \Psi^0_n |n\ra$.
 The Schr\"{o}dinger equation,
\begin{align}\label{S-eq}
 i\frac{d}{dt} |\Psi(t)\ra = H |\Psi(t)\ra
\end{align}
 with $\hbar = 1$, predicts the time evolution of
 $ |\Psi(t)\ra = \sum_n \Psi^0_n e^{-iE_nt } |n\ra$.
 In the branch formulation, it is written as follows,
\begin{equation}
    |\Psi(t)\ra = \sum_{\alpha} \ket{\alpha}  \ket{\phi_\alpha ^\E(t)},
\end{equation}
 where the \textit{environmental branch} $|\phi_\alpha^\E(t)\ra$, as a vector in the
 space of $\HH_\E$, is given by
\begin{align}\label{}
 & |\phi_\alpha^\E(t)\ra = \la \alpha |\Psi \ra
 = \sum_n \Psi^0_n e^{-iE_nt } \la \alpha |n\ra.
\end{align}
 A direct computation from Eq.(\ref{S-eq}) gives that
\begin{align}\label{}
     i \frac{d}{dt} \ket{\phi_\alpha ^\E  }
     = (e^S_\alpha + H^\E) \ket{\phi_\alpha ^\E  } +
     \sum_{\gamma }  H^I_{\alpha \gamma} \ket{\phi_\gamma ^\E  }, \label{dphi-dt}
\end{align}
where $H^I_{\alpha \beta} \equiv \la \alpha |H^I | \beta \ra$ as an environmental operator is written as
\begin{align}\label{}
 H^I_{\alpha \beta}  =  H^{IS}_{\alpha \beta} \, H^{I\E}.
\end{align}

Taking the partial trace of $\ket{\Psi(t)}\bra{\Psi(t)}$ over the environmental degree of freedom
yields the RDM, $\rho^S(t) = \tr_\E (\ket{\Psi(t)}\bra{\Psi(t)})$.
It is straightforward to verify the following well known relation, 
\begin{equation}\label{rhoab-phiba}
 \rho^S_{\alpha \beta}(t) = \la \phi_\beta^\E(t)|\phi_\alpha^\E(t) \ra,
\end{equation}
 showing that the RDM elements can be simply written as inner products of the environmental branches.
 Making use of Eq.(\ref{dphi-dt}), it is straightforward to get the following
 evolution equation of the elements,
\begin{align}\label{}
    & \frac{d \rho^S_{\alpha \beta}}{d t}
     = i (e^S_\beta - e^S_\alpha) \rho^S_{\alpha \beta}
     + i \sum_{\gamma } (  H^{I\E}_{\phi, \alpha \gamma} H^{IS}_{\gamma  \beta }
     - H^{IS}_{\alpha \gamma  } H^{I\E}_{\phi, \gamma \beta} ),
\label{drho-dt-Gammat}
\end{align}
 where $H^{I\E}_{\phi, \alpha \beta}$ is a c-number defined by
\begin{align}\label{HIE-phiab}
 H^{I\E}_{\phi, \alpha \beta}(t) := \la \phi_\beta^\E(t)| H^{I\E} |\phi_\alpha ^\E(t)\ra.
\end{align}

 Expanding the environmental branches in the environmental eigenstates, i.e.,
\begin{align}\label{phi-expan-C}
 |\phi^\E_\alpha(t)\ra = \sum_{i } C_{\alpha i}(t) |i\ra,
\end{align}
 one gets that
\begin{align}\label{rho-CC}
 \rho^S_{\alpha \beta}(t)
 =  \sum_{i }  C^*_{\beta i}(t) C_{\alpha i}(t).
\end{align}
 From Eq.(\ref{drho-dt-Gammat}), one sees that quantities like $H^{I\E}_{\phi, \alpha \beta}$
 play a key role in the evolution of the RDM.
 Inserting Eq.(\ref{phi-expan-C}) into Eq.(\ref{HIE-phiab}), one gets that
\begin{align}
 &   H^{I\E}_{\phi, \alpha \beta} = \sum_{i, j  } C^*_{\beta j}(t) C_{\alpha i}(t) H^{I\E}_{ji}.
\label{h2IE-sum}
\end{align}
 This quantity behaves in a complicated way with time, particularly due to
 correlations between the branch-expansion components [$C_{\alpha i}(t)$ and $C_{\beta j}(t)$]
 and the matrix elements $H^{I\E}_{ji}$.
 Such correlations originate from the role played by $H^{I\E}$
 in the time evolution of the environmental branches,
 as a part of the interaction Hamiltonian.
 \footnote{
 This point is seen clearly in Eq.(\ref{Phi(t)-exp}) to be derived later.}

\subsection{Preliminary analysis based on ETH ansatz}
\label{HIE-under-ETH}

 In this section, we discuss implications of the ETH ansatz  on
 properties of $H^{I\E}_{\phi, \alpha \beta}$ in Eq.(\ref{h2IE-sum}).

 We use $\Gamma$ to indicate the environmental energy window,
 within which all significant environmental branches lie (for all times concerned).
 Its width is denoted by $\Delta_{\Gamma}$,  its center by $e^\Gamma_{c}$,
 and the number of levels in it by $M_{\Gamma}$.
 The energy window for one environmental branch $|\phi_\alpha^\E(t)\ra$ 
 is indicated by $\Gamma_\alpha$, 
 with a number of $M_{\Gamma_\alpha}$ levels.
 Clearly, the sum of $\Gamma_\alpha$ gives $\Gamma$, namely,
 $\Gamma = \cup_\alpha \Gamma_\alpha$.

 We are to consider interactions that are not very weak.
 This implies that, usually, $\Delta_{\Gamma_\alpha}$ are not very small and,
 due to the many-body feature of the environment, $M_{\Gamma_\alpha}$
 are very large numbers. 
 Then, for a small system $S$,  in a scaling analysis 
 with respect to $M_\Gamma$, the difference
 between $\Gamma$ and $\Gamma_\alpha$ is negligible.
 Thus, e.g., due to normalization of the total state, 
 the expansion coefficients of the environmental branches 
 in Eq.(\ref{phi-expan-C}) scale as $C_{\alpha i} \sim M_{\Gamma}^{-1/2}$.

 One important technique, which is to be frequently employed in what follows,
 is to divide quantities like $H^{I\E}_{\phi, \alpha \beta}$
 in Eq.(\ref{h2IE-sum}) into four parts
 according to the ETH-ansatz, say, for $H^{I\E}_{ij}$ 
 given by the rhs of Eq.(\ref{ETH}) with $O=H^{I\E}$.
 To do this, 
 we use $h^{I\E}(e)$ to indicate the diagonal function for $H^{I\E}$ on the rhs of Eq.(\ref{ETH})
 and use $h_{0}^{I\E}$ to indicate the value of $h^{I\E}(e)$ at the center
 of $\Gamma$, namely at $e=e^\Gamma_{c}$.
 This division is to be labelled by $l$  of $l=1,2,3,4$ and the result for 
 $H^{I\E}_{\phi, \alpha \beta}$ is to be
 indiciated by $H^{I\E (l)}_{\phi, \alpha \beta}$, written as
\begin{align}
 &  H^{I\E}_{\phi, \alpha \beta} = \sum_{l=1}^4 H^{I\E (l)}_{\phi, \alpha \beta}.
\label{HIE-phiab=sum-l}
\end{align}
 Explicit meanings of the label $l$ for the corresponding contributions
 are as follows:
\begin{subequations}\label{HIE-l=1,4}
\begin{align} \notag
 & \text{$l=1$: from $h_{0}^{I\E}$, the value of the diagonal  function}
 \\ & \quad \text{  $h^{I\E}(e)$ taken at the point of $e=e^\Gamma_{c}$;}
 \\  & \text{$l=2$: from deviation of $h^{I\E}(e)$ from $h_{0}^{I\E}$; }
 \\  & \text{$l=3$: from fluctuations of diagonal elements;}
 \\  & \text{$l=4$: from offdiagonal elements.}
\end{align}
\end{subequations}
 (See below for explicit expressions of $H^{I\E (l)}_{\phi, \alpha \beta}$.)

 Firstly, let us discuss the case of $l=1$.
 As the contribution from $h_{0}^{I\E}$ to the rhs of Eq.(\ref{h2IE-sum}), it is written as
\begin{align}
 &   H^{I\E (1)}_{\phi, \alpha \beta} = \sum_{i } C^*_{\beta i}(t) C_{\alpha i}(t) h_{0}^{I\E} = h_{0}^{I\E} \rho^S_{\alpha \beta }(t),
\label{HIE-phiab-(1)}
\end{align}
 where Eq.(\ref{rho-CC}) has been used.
 It is seen that $H^{I\E (1)}_{\phi, \alpha \beta}$ is directly related to the system's RDM element.
 This is in fact a generic phenomenon; that is, with the
 ETH ansatz applied to $\la \phi_\beta^\E(t)|O |\phi_\alpha ^\E(t)\ra$
 for an arbitrary observable $O$, the contribution of $l=1$ is proportional to $\rho^S_{\alpha \beta }(t)$.

 One observation is that the term $H^{I\E (1)}_{\phi, \alpha \beta}$ may be absorbed
 into the self-Hamiltonian of the system $S$ by a renormalization.
 Generically, one may write the total Hamiltonian in a renormalized way as follows,
\begin{align}\label{H-renormalize}
 H = H^S_{\rm rn} + H^\E + H^I_{\rm rn},
\end{align}
 where
\begin{align}\label{}
 H^S_{\rm rn} = H^S + O^S, \quad H^I_{\rm rn} = H^I - O^S.
\end{align}
 Here, $O^S$ is an operator acting on the Hilbert space of the central system $S$, which represents certain averaged impact of the system-environment 
 interaction on the system $S$.
 For the problem at hand, we take
\begin{align}\notag
 & O^S = h_{0}^{I\E} H^{IS} 
 \\ & \Rightarrow 
 H^S_{\rm rn} = H^S + h_{0}^{I\E} H^{IS}, \ 
 H^I_{\rm rn} = H^{IS} (H^{I\E} - h_0^{I\E}). \label{OS-2}
\end{align}
 It is straightforward to check that 
 $H^{I\E (1)}_{{\rm rn} \phi, \alpha \beta} =0$ for the renormalized Hamiltonian,
 according to a definition similar to that in Eq.(\ref{HIE-phiab-(1)}).
 Below, we assume that this renormalization produce has been taken,
 \footnote{Clearly, with this procedure, the basis states $|\alpha\ra$ should be
 eigenstates of $H^S_{\rm rn}$.}
 which is effectively equivalent to the case of $h_0^{I\E} = 0$,
 and for brevity we omit the subscript ``rn''.

 Secondly, we discuss the case of $l=2$ for
 the contribution of deviation of $h^{I\E}(e)$ from $h_0^{I\E}$, which is written as
\begin{align}
 &  H^{I\E (2)}_{\phi, \alpha \beta} = \sum_{i}  C^*_{\beta i}(t) C_{\alpha i}(t)
 \big( h^{I\E}(e^\E_i)-h_0^{I\E} \big).
\label{HIE-phiab-(2)}
\end{align}
 According to the ETH ansatz, $h^{I\E}(e)$ is a slowly varying function.
 Hence, for an energy shell $\Gamma$ not wide,
 linear approximation should be useful for $h^{I\E}(e)$ within $\Gamma$, namely,
 $h^{I\E}(e) \simeq h_0^{I\E} + (e - e^\Gamma_{c}) {h'_0}^{I\E}$, where ${h'_0}^{I\E}$ indicates
 the derivative of the diagonal function at the point $e^\Gamma_{c}$.
 This gives that
\begin{align}\label{HIE(2)-CCh'}
 H^{I\E (2)}_{\phi, \alpha \beta} \simeq  {h'_0}^{I\E} \sum_{i \in \Gamma}  C^*_{\beta i}(t) C_{\alpha i}(t)  (e^\E_i - e^\Gamma_{c}).
\end{align}
 Clearly, unlike $H^{I\E (1)}_{\phi, \alpha \beta}$,
 generically $H^{I\E (2)}_{\phi, \alpha \beta}$
 can not be written as a function of the RDM.

 One may further note that $h^{I\E}(e)$ may appear as a function of ${e}/{N}$ in some situations
 \cite{KIH-pre14,Rigol-prb20,Rigol-AiP16,pra22-PB-ETH,pre25-ETHsc}.
 In this case, the slope ${h'_0}^{I\E}$ is
 reversely proportional to the particle number $N$ 
 and goes to zero in the limit of large $N$.
 As a result, $H^{I\E (2)}_{\phi, \alpha \beta}$ is negligible for a sufficiently large $N$.

 However, it is unclear whether the slope ${h'_0}^{I\E}$ could 
 be very small in a generic situation.
 As discussed previously, one purpose of this paper is to study situations, in which the RDM 
 could be approximately enough for determining its time evolution.
 For this reason, below, we are to assume that the slope ${h'_0}^{I\E}$ is sufficiently small,
 such that the terms $H^{I\E (2)}_{\phi, \alpha \beta}$ of $l=2$ are negligible.
\footnote{When terms like $H^{I\E (2)}_{\phi, \alpha \beta}$ are nonnegligible,
 their contributions to the rhs of Eq.(\ref{drho-dt-Gammat}) may induce effects
 sometimes under the name of nonMarkovian.}

 Thirdly, we discuss the case of $l=3$,
 which is for fluctuations of the diagonal elements $H^{I\E}_{ii}$.
 This part is written as
\begin{align}
 &  H^{I\E (3)}_{\phi, \alpha \beta} = \sum_{i}  C^*_{\beta i}(t) C_{\alpha i}(t)
 \frac{1}{\sqrt{\rho_{\rm dos}^\E(e^0)}} f(e^0,\omega) r_{ii}.
\label{HIE-phiab-(3)}
\end{align}
 Since the major branch components which lie within $\Gamma$
 scale as $C_{\alpha(\beta) i} \sim M_\Gamma^{-1/2}$ ,
 the rhs of Eq.(\ref{HIE-phiab-(3)}) shows that 
 $H^{I\E (3)}_{\phi, \alpha \beta}$ at most scales as
 $(\rho_{\rm dos}^\E(e^0))^{-1/2}$,
 which behaves as $M_\Gamma^{-1/2}$.
 Hence, this part is quite small for a large environment and is usually negligible.

 Finally, we discuss the case of $l=4$ for the contribution from
 offdiagonal elements $H^{I\E}_{ij}$ of $i\ne j$.
It is written as
\begin{align}\notag
    H^{I\E (4)}_{\phi, \alpha \beta} & = \sum_{i \ne j}
 \la \phi_\beta^\E(t)| j\ra \la j| H^{I\E} |i\ra \la i |\phi_\alpha ^\E(t)\ra
 \\ & = \sum_{i \ne j } C^*_{\beta j}(t) C_{\alpha i}(t) H^{I\E}_{ji}.
\label{HIE-phiab-(4)}
\end{align}
 Since ${C_{\alpha i}} \sim M_\Gamma^{-1/2}$ within $\Gamma$
 and $H^{I\E}_{ji}$ scales as $M_\Gamma^{-1/2}$ according to the ETH ansatz,
 one sees that $C^*_{\gamma i}(t) C_{\alpha j}(t) H^{I\E}_{ji} \sim M_{\Gamma}^{-3/2}$.
 Generically, one may use $(M_{\Gamma})^{\lambda}$ with a parameter $\lambda$ to represent
 the scaling contribution from the summation over $i$ and $j$ on the rhs of Eq.(\ref{HIE-phiab-(4)}).
 Then, one has
\begin{align}\label{HIE4-M-scale}
  & H^{I\E (4)}_{\phi, \alpha \beta}(t)  \sim M_\Gamma^{\lambda-1.5} \sim (\rho_{\rm dos}^\E)^{\lambda-1.5}.
\end{align}

 To summarize the above discussions, 
 $h_0^{I\E}$ can be set zero by a renormalization of 
 the Hamiltonian, which eliminates $H^{I\E (1)}_{\phi, \alpha \beta}$;
 the slope ${h'_0}^{I\E}$ is assumed sufficiently small,
 which implies smallness of $H^{I\E (2)}_{\phi, \alpha \beta}$;
 and $H^{I\E (3)}_{\phi, \alpha \beta}$ is small due to its scaling behavior.
 Thus,  one finds that
\begin{align}\label{HIE-simeq-HIE4}
 H^{I\E }_{\phi, \alpha \beta} \simeq H^{I\E (4)}_{\phi, \alpha \beta}.
\end{align}

 Below, we argue that $\lambda$ should be equal to $1.5$ in the limit of large $N$,
 if the RDM undergoes a nonnegligible evolution beside phase changes.
 In fact, in such a case, the second part on the rhs of Eq.(\ref{drho-dt-Gammat})
 should differ notably from zero, which implies that
 at least some of $H^{I\E }_{\phi, \alpha \gamma}$ or $H^{I\E }_{\phi, \gamma \beta}$
 should be notably different from zero.
 Then, since contributions from $l=1,2,3$ are negligible,
 at least some of $H^{I\E (4)}_{\phi, \alpha \gamma}$ or $H^{I\E (4)}_{\phi, \gamma \beta}$
 should be not very small.
 According to the scaling behavior in Eq.(\ref{HIE4-M-scale}), this requires that
 $\lambda =1.5$ in the limit of $N\to \infty$ with $\rho_{\rm dos}^\E \to \infty$.

 The value of $\lambda =1.5$ implies certain phase correlation 
 among the three quantities of $C^*_{\beta j}(t)$, $C_{\alpha i}(t)$, 
 and $H^{I\E}_{ji}$; in other words, 
 between the environmental branches and the operator $H^{I\E}$ in the interaction Hamiltonian.
 In fact, if the three quantities could possess purely random phases,
 one would have $\lambda =1$.
 While, note that phases of the three quantities can not fully compensate,
 which would imply $\lambda=2$.
\footnote{See Sec.\ref{sect-HIE4-lambda} for further discussions on
 the correlations and the value of $\lambda$.}

 Even with the simplification in Eq.(\ref{HIE-simeq-HIE4}),
 it is still not an easy task to compute variation of the RDM 
 from Eq.(\ref{drho-dt-Gammat}).
 This is particularly due to a two-fold feature of the quantity 
 $H^{I\E (4)}_{\phi, \alpha \beta}$:
 Although it may possess a very small mean value,
 the above discussed three quantities in it possess certain phase correlations.

\section{Finite-time evolution of RDM}\label{sect-finite-t-evolve}

 In this section, we first derive an evolution equation for environmental branches (Sec.\ref{sect-formal-solution});
 then, making use of a formal solution to the equation,
 study a finite-time evolution of the RDM under the ETH ansatz 
 (Secs.\ref{sect-evolv-t0-t} and \ref{sect-evolv-t0-t-under-ETH});
 and, finally, discuss some properties of $H^{I\E (4)}_{\phi, \alpha \beta}$ (Sec.\ref{sect-HIE4-lambda}).

\subsection{Formal solution to branch evolution}\label{sect-formal-solution}

 It is easy to see that, under nondissipative interactions 
 for which $H^{IS}_{\alpha  \gamma } =0$ for all $\alpha \ne \gamma$,
 Eq.(\ref{dphi-dt}) reduces to a Schr\"{o}dinger-type equation under certain effective Hamiltonian
 and the environmental branches $|\phi^\E_\alpha(t)\ra$ undergo a Schr\"{o}dinger-type evolution.
 Below, we show that this property is generalizable to a generic dissipative interaction.

 To study branch evolution in a generic case, let us write Eq.(\ref{dphi-dt}) in the following form:
\begin{align}\label{dphi-alpha-dt}
     i \frac{d}{dt} \ket{\phi_\alpha ^\E  } = \sum_{\gamma} \M_{\alpha \gamma} \ket{\phi_\gamma ^\E  },
\end{align}
where $\M$ is a $d_S \times d_S$ matrix, defined by
\begin{align}\label{}
 & (\M )_{\alpha \gamma} = (e^S_\alpha + H^\E) \delta_{\alpha \gamma} +  H^{I\E} H^{IS}_{\alpha \gamma}.
\end{align}
 The expression in Eq.(\ref{dphi-alpha-dt}) suggests
 that one may introduce an auxiliary notation with respect to the label $\alpha$
 and write the environmental branches in a matrix form, 
 which is to be refereed to as \textit{$\alpha$-matrix} and indicated as $[ \cdot ]$.
 Specifically,  the $d_S$ kets of the branches are written in the following matrix form,
 \footnote{Although the $\alpha$-matrix notation is related to the label 
 $\alpha$ of the eigenbasis of the system $S$, this notation is different from
 the eigenbasis-representation of the system $S$,
 as seen clearly from Eq.(\ref{[phi-E]}).
}
\begin{align}\label{[phi-E]}
  &[|\phi^\E \ra] := \left(                 \begin{array}{c}                    |\phi_1 ^\E\ra \\
                   |\phi_2 ^\E\ra \\                    \vdots \\                    |\phi_{d_S} ^\E\ra \\
                 \end{array}
               \right),
\end{align}
 and, similarly, the bras are  written as
\begin{align}\label{}
 & [\la \phi^\E |] \equiv \left(\la \phi_1 ^\E|,\;
                   \la \phi_2 ^\E| ,\;                    \cdots ,\;                    \la \phi_{d_S} ^\E| \right).
\end{align}
 In the notation of $\alpha$-matrix, it is easy to check that the RDM has the following matrix expression,
\begin{subequations}\label{}
\begin{align}\label{rho-tr-PhiE}
 & [\rho^S] = \tr_\E \Big( [|\phi^\E\ra] [ \la \phi^\E|] \Big),
 \\ & \label{rho-tr-PhiE-ab}
 \rho^S_{\alpha \beta}(t) = \big[ \la \phi^\E(t)| \big]_\beta  \big[|\phi^\E(t)\ra \big]_\alpha ,
\end{align}
\end{subequations}
 where $\big[|\phi^\E(t)\ra \big]_\alpha$ represents the $\alpha$th row of $\big[|\phi^\E(t)\ra \big]$
 and $\big[ \la \phi^\E(t)| \big]_\beta$ the $\beta$th column of $\big[ \la \phi^\E(t)| \big]$.

 Then, Eq.(\ref{dphi-alpha-dt}) is written in a compact form,
\begin{align}\label{}
     i \frac{d}{dt} [|\phi^\E\ra ] = {\M} [|\phi^\E\ra],     \label{dPhi-dt}
\end{align}
 where
\begin{align}\label{M-matrix}
 & \M = [H^S] +  H^\E [I] +  H^{I\E}[H^{IS}].
\end{align}
 Here, $[H^{S}]$ and $[H^{IS}]$, as the $\alpha$-matrices for $H^S$ and $H^{IS}$, respectively,
 are equal to the matrices of $H^{S}_{\alpha \beta}$ and $H^{IS}_{\alpha \beta}$;
 and $[I]$ indicates the $d_S \times d_S$ unit matrix.
 Equation (\ref{dPhi-dt}) has the following formal solution for the 
 environmental branches' evolution,
\begin{align}\label{Phi(t)-exp}
    [|\phi^\E(t) \ra]  = e^{-i \M (t-t_0)} [|\phi^\E(t_0)\ra].
\end{align}

Furthermore, in the $\alpha$-matrix notation, the evolution equation of the
 RDM in Eq.(\ref{drho-dt-Gammat}) is written in a concise way,
\begin{align}\label{}
    & i\frac{d [\rho^S]}{d t}
     =  \Big[ [H^S], [\rho^S] \Big] + \Big[ [H^{IS}], [H^{I\E}_{\phi}] \Big],
\label{drho-dt-matrix}
\end{align}
 where $[H^{I\E}_{\phi}]$ indicates an $\alpha$-matrix whose elements are
 $[H^{I\E}_{\phi}]_{\alpha \beta} = H^{I\E}_{\phi, \alpha \beta}$.
 If $[H^{I\E}_{\phi}]$ could be (approximately and under some condition) written as 
 a linear function of $[\rho^S]$, 
 then, a master equation would be
 obtainable from Eq.(\ref{drho-dt-matrix}),
 though this is not a simple task.

\subsection{Formal expression for finite-time evolution}\label{sect-evolv-t0-t}

 In this section, based on the above obtained formal solution to the environmental-branch evolution [Eq.(\ref{Phi(t)-exp})], 
 a formal expression is derived for the  
 evolution of the RDM within a finite time period.

 To get an explicit expression for the RDM elements from Eq.(\ref{rho-tr-PhiE-ab}),
 let us write explicitly the component $\big[|\phi^\E(t)\ra \big]_\alpha $
 according to  Eq.(\ref{Phi(t)-exp}), namely, 
\begin{align} \notag
  \big[|\phi^\E(t)\ra \big]_\alpha & =
 ( e^{-i \M (t-t_0)}\big[ |\phi^\E(t_0)\ra \big] )_\alpha
 \\ & = \sum_{n=0}^{\infty} \frac{(-i (t-t_0))^n}{n!} \left(\M^n \big[ |\phi^\E(t_0)\ra \big] \right)_\alpha.
\label{|Phi>-exp}
\end{align}
 Similarly, the bra component is written as
\begin{align} \notag  \big[\la \phi^\E(t)| \big]_\beta & =
 (\big[ \la \phi^\E(t_0)| \big] e^{i \M (t-t_0)} )_\beta
 \\ & = \sum_{n=0}^{\infty} \frac{(i (t-t_0))^n}{n!} \left(\big[ \la \phi^\E(t_0)| \big] \M^n \right)_\beta.
 \label{<Phi|-exp}
\end{align}
 Substituting the above expansions into Eq.(\ref{rho-tr-PhiE-ab}), one gets
\begin{align}\notag
 \rho^S_{\alpha \beta}(t) & = \sum_{n,m=0}^{\infty} (-1)^m \frac{(i (t-t_0))^{n+m}}{n! m!}
 \\ & \times \Big( \big[ \la \phi^\E(t_0)| \big] \M^n \Big)_\beta
  \Big( \M^m \big[ |\phi^\E(t_0)\ra \big] \Big)_\alpha.
  \label{rho-phi1}
\end{align}
 With $k=n+m$, the above expansion is rewritten as
\begin{align}
 & \rho^S_{\alpha \beta}(t)  - \rho^S_{\alpha \beta}(t_{0})
  =  \sum_{k=1}^{\infty} G^{(k)}_{\alpha \beta}(t_{0}) (t-t_0)^{k},
 \label{drho-G-t^k}
\end{align}
 where Eq.(\ref{rhoab-phiba}) has been used for $k=0$ and
 $G^{(k)}_{\alpha \beta}(t)$ of $k \ge 1$ is defined as follows,
\begin{align} \notag
 G^{(k)}_{\alpha \beta}(t) & = i^k \sum_{n=0}^{k} \frac{ (-1)^{k-n}}{n! (k-n)!}
 \\ & \times \Big( \big[ \la \phi^\E(t)| \big] \M^n \Big)_\beta  
 \Big(\M^{k-n} \big[ |\phi^\E(t)\ra \big] \Big)_\alpha.
\label{Gkt0=sum-phi}
\end{align}

 Let us expand $\Big(\M^{n} \big[ |\phi^\E(t_0)\ra \big] \Big)_\alpha$
 with $\M$ given in Eq.(\ref{M-matrix}).
 Formally, it is written as 
\begin{align}\label{Mm-phi-expand}
  \Big(\M^{n} \big[ |\phi^\E(t_0)\ra \big] \Big)_\alpha
  = \sum_{r,\gamma} u_{\alpha  \gamma}^{n,r} X_{\alpha  \gamma}^{n,r} |\phi_{\gamma}^\E(t_0)\ra,
\end{align}
 where $u_{\alpha  \gamma}^{n,r}$ and $X_{\alpha  \gamma}^{n,r}$ 
 to represent coefficients and environmental operators that appear in
 the expansion, respectively,  with a label $r$ introduced for the expansion.
 Making use of Eq.(\ref{Mm-phi-expand}) and its bra form,
 the second line in Eq.(\ref{Gkt0=sum-phi}) is written as 
\begin{align}\label{u*u-XX-phi}
  \sum_{r,\gamma,r',\gamma'}   u_{\beta  \gamma'}^{n,r' *}   u_{\alpha  \gamma}^{k-n,r}
   \la \phi_{\gamma'}^\E(t_0)|X_{\beta  \gamma'}^{n,r' } 
   X_{\alpha  \gamma}^{k-n,r} |\phi_{\gamma}^\E(t_0)\ra.
\end{align}

 Some of the operator products $X_{\beta  \gamma'}^{n,r' } X_{\alpha  \gamma}^{k-n,r}$ 
 in (\ref{u*u-XX-phi}) may be identical
 and we use $Y_{\eta}$ to indicate independent ones among them,
 with a label $\eta$ introduced for this purpose. 
 Then, inserting the result obtained into Eq.(\ref{Gkt0=sum-phi}),
 one gets the following expression of $G^{(k)}_{\alpha \beta}(t_{0})$,
\begin{align}\label{Gk-sum-xYphi}
 G^{(k)}_{\alpha \beta}(t_{0}) =  \sum_{\eta}  G^{(k)}_{\alpha \beta, \eta}(t_{0}),
\end{align}
 where 
\begin{align}\label{Gketa-sum-xYphi}
 G^{(k)}_{\alpha \beta, \eta}(t) =  \sum_{\gamma,\gamma'}  
 x_{\alpha \beta, \eta}^{(k)  \gamma \gamma'}
 \la \phi_{\gamma'}^\E(t)|  Y_{\eta} |\phi_{\gamma}^\E(t)\ra;
\end{align}
 here, $x_{\alpha \beta, \eta}^{(k)  \gamma \gamma'}$ indicate
 the coefficients that appear in this expansion.
 Note that the values of $x_{\alpha \beta, \eta}^{(k)  \gamma \gamma'}$
 are determined by the Hamiltonian, independent of the initial state of 
 the total system, and, 
 hence, can be regarded as parameters, though usually
 they not independent of each other.

\subsection{Finite-time evolution under ETH}\label{sect-evolv-t0-t-under-ETH}

 In this section, with the help of the ETH ansatz,
 we discuss evolution of the RDM within a finite time period. 

 It can be shown that,  
 \footnote{See Appendix \ref{app-argu-Gk} for detailed discussion on this property, where a method of computing the operators $Y_{\eta}$ is given.
 And, some examples  of the operators $Y_{\eta}$ are to be given in 
 Eq.(\ref{Yk-01}).
 }
 due to locality of the interaction,
 those operators $Y_{\eta}$, which appear in the expansion at $k$ much 
 smaller than the particle number $N$, are either the environmental identity operator $I^\E$ 
 or some local operator, such as $H^{I\E}$ and $(H^{I\E})^2$.
 Note that the identity operator $I^\E$  can be regarded as satisfying the ETH ansatz in Eq.(\ref{ETH}),
 with vanishing fluctuation function of $f(e^0,\omega)=0$;
 meanwhile, as mentioned previously, local operators are usually believed to satisfy the ETH ansatz.
 We use $k_{\rm ETH}$ to indicate an upper bound of $k$,
 such that all the operators $Y_{\eta}$ corresponding to 
 $k \le k_{\rm ETH}$ obey the ETH ansatz.
 Following arguments given in Appendix \ref{app-argu-Gk},
 one finds that $k_{\rm ETH}$ may go to infinity in the thermodynamic limit of $N\to \infty$.

 For $k \le k_{\rm ETH}$ with $Y_{\eta}$ satisfying the ETH ansatz,
 the quantity $\la \phi_{\beta}^\E(t)|  Y_{\eta} |\phi_{\alpha}^\E(t)\ra$,
 which is to be indicated as $Y_{\eta \phi, \alpha \beta}(t)$, 
 can be treated in a way similar to that
 given previously for $H^{I\E}_{\phi, \alpha \beta}$ in Sec.\ref{HIE-under-ETH}.
 In particular, it is divided into four parts also labelled with $l$, namely, 
\begin{align}\label{Yphi}
  Y_{\eta \phi, \alpha \beta}(t) := 
 \la \phi_{\beta}^\E(t)|  Y_{\eta} |\phi_{\alpha}^\E(t)\ra
  \equiv \sum_{l=1}^4  Y^{(l)}_{\eta \phi, \alpha \beta},
\end{align}
 where the meaning of $l$ is similar to that given in Eq.(\ref{HIE-l=1,4}),
 but with $H^{I\E}$ replaced by $Y_{\eta}$.
 This gives a further division of $G^{(k)}_{\alpha \beta,\eta}(t_{0})$,
\begin{align}\label{Omega-4}
 G^{(k)}_{\alpha \beta, \eta}(t_{0}) = \sum_{l=1}^4 G^{(k,l)}_{\alpha \beta, \eta}(t_{0}),
\end{align}
 where
\begin{align}\label{Gketa-l-sum-xYphi}
 G^{(k,l)}_{\alpha \beta, \eta}(t) =  \sum_{\gamma,\gamma'}  
 x_{\alpha \beta, \eta}^{(k)  \gamma \gamma'} 
 Y^{(l)}_{\eta \phi, \gamma \gamma'}(t).
\end{align}

 We use $y_{\eta}(e)$ to indicate the diagonal function for $Y_{\eta}$
 in the ETH ansatz and use $y_{\eta 0}$ to indicate its value
 at the center $e^\Gamma_{c}$ of the window $\Gamma$, 
 $y_{\eta 0} = y_{\eta}(e^\Gamma_{c})$.
 Thus, for $l=1$, similar to Eq.(\ref{HIE-phiab-(1)}),
 one has 
\begin{align}\label{Yketa-phiab-(1)}
  Y^{(1)}_{\eta \phi, \gamma \gamma'}(t) = \sum_{i } C^*_{\gamma' i}(t) C_{\gamma i}(t) y_{\eta 0} 
 = y_{\eta 0} \rho^S_{\gamma \gamma'}(t),
\end{align} 
and, as a consequence,
\begin{align}\label{Gk-l1-sum}
 G^{(k,1)}_{\alpha \beta, \eta}(t) = \sum_{\eta,\gamma,\gamma'} y_{\eta 0}
 x_{\alpha \beta, \eta}^{(k)  \gamma \gamma'}  \rho^S_{\gamma \gamma'}(t).
\end{align}

 For $l=2$, the quantity $Y^{(l)}_{\eta \phi, \alpha \beta}$ comes
 from deviation of $y_{\eta}(e)$ from $y_{\eta 0}$.
 Similar to $H^{I\E (2)}_{\phi, \alpha \beta}$ in Eq.(\ref{HIE-phiab-(2)}),
 it is written as 
\begin{align}
 &  Y^{(2)}_{\eta\phi, \alpha \beta}(t) 
 = \sum_{i}  C^*_{\beta i}(t) C_{\alpha i}(t)
 \Big( y_{\eta}(e^\E_i)-y_{\eta 0} \Big).
\label{Yketa-phiab-(2)}
\end{align}
 Following arguments similar to those given previously for $H^{I\E}$
 (as a special case of $Y_{\eta}$),
 one sees that $Y^{(2)}_{\eta\phi, \alpha \beta}(t)$
 is written in a form like the rhs of Eq.(\ref{HIE(2)-CCh'}), but with ${h'_0}^{I\E}$ replaced by
 the slope $y'_{\eta 0}$ of the function $y_{\eta}(e)$,
 which is also taken at the center of $\Gamma$,  i.e., 
\begin{align}\label{Yketa(2)-CCh'}
 Y^{(2)}_{\eta\phi, \alpha \beta}(t) \simeq  y'_{\eta 0} \sum_{i}
 C^*_{\beta i}(t) C_{\alpha i}(t)  (e^\E_i - e^\Gamma_{c}).
\end{align}
 Like the case of $H^{I\E (2)}_{\phi, \alpha \beta}$ discussed previously,
 we assume that $y'_{\eta 0}$ are sufficiently small, 
 such that $G^{(k,2)}_{\alpha \beta, \eta}(t_{0})$ are negligible.

 For $l=3$ for the contribution from fluctuations of diagonal elements,
 following arguments similar to those given previously around Eq.(\ref{HIE-phiab-(3)}),
 one sees that $G^{(k,3)}_{\alpha \beta, \eta}(t_{0})$ 
 should be negligible.

 For $l=4$ with offdiagonal elements, $G^{(k,4)}_{\alpha \beta, \eta}(t_{0})$
 should have the same scaling behavior
 as that shown in Eq.(\ref{HIE4-M-scale}), namely, $M_\Gamma^{\lambda-1.5}$.
 One should note that the previous argument for $\lambda =1.5$ 
 does not include the initial time.
 In fact, for an initial environmental state that possesses
 no correlation with the operator $H^{I\E}$, it is possible for $\lambda$
 to be equal to $1$.

 To be specific, one may consider an initial condition in which the environment
 lies in a thermal state (or a typical state within the related energy shell).
 In this case, one has $\lambda =1$ and
 $G^{(k,4)}_{\alpha \beta, \eta}(t_{0}) \sim M_\Gamma^{-1/2}$, being negligible.
 Then, one gets that
\begin{align}\label{Gk=Gk1-t0}
 G^{(k)}_{\alpha \beta, \eta}(t_{0}) \simeq G^{(k,1)}_{\alpha \beta, \eta}(t_{0}) \quad \text{for $k \le k_{\rm ETH}$}.
\end{align}

 For a given time $t$, we use $k_{\rm tru}(t)$ to indicate a value of $k$,
 at which the rhs of Eq.(\ref{drho-G-t^k}) may be effectively truncated;
 and below, for brevity, we usually do not write explicitly the dependence of $k_{\rm tru}$ on $t$.
 Suppose that $(t-t_0)$ is not very large, such that $k_{\rm tru} \le k_{\rm ETH}$.
 Then, substituting Eq.(\ref{Gk-l1-sum}) with Eq.(\ref{Gk=Gk1-t0}) into Eq.(\ref{drho-G-t^k}), one gets that
\begin{align}\notag
 & \rho^S_{\alpha \beta}(t)   \simeq
 \rho^S_{\alpha \beta}(t_{0}) + \sum_{k=1}^{k_{\rm tru}} G^{(k)}_{\alpha \beta}(t_{0}) (t-t_0)^{k}
 \\ & \simeq \rho^S_{\alpha \beta}(t_{0})
 + \sum_{k=1}^{k_{\rm tru}} \sum_{\eta,\gamma,\gamma'} y_{\eta 0}
 x_{\alpha \beta, \eta}^{(k)  \gamma \gamma'}
  \rho^S_{\gamma \gamma'}(t_0) (t-t_0)^{k} .
 \label{rho(t)-G-t^k-t0}
\end{align}

 To summarize, the RDM $\rho^S$ at a time $t$
 is approximately computable from the initial RDM $\rho^S(t_0)$,
 if the following conditions are fulfilled.
 That is, (i) the environment satisfies the ETH ansatz;
 (ii) all the slopes $y'_{\eta 0}$ of $k \le k_{\rm tru}(t)$ are sufficiently small,
 such that those contributions with $l=2$ can be neglected;
 and (iii) initially the environment lies in a thermal state.
 However, practical computation of $\rho^S(t)$ by making use of 
 Eq.(\ref{rho(t)-G-t^k-t0}) usually becomes quite difficult 
 when $(t-t_0)$ is not small, which may require a large value of $k_{\rm tru}(t)$.

\subsection{Some properties of $H^{I\E (4)}_{\phi, \alpha \beta}$}
\label{sect-HIE4-lambda}

 In this section, making use of the above derived formal expression 
 for the environmental-branch evolution in Eq.(\ref{Phi(t)-exp}), we discuss two properties of 
 the quantity $H^{I\E (4)}_{\phi, \alpha \beta}(t)$ from the dynamic perspective.
 One is about value of the scaling parameter $\lambda$ in Eq.(\ref{HIE4-M-scale}).
 The other is about decay of phase correlation of $H^{I\E (4)}_{\phi, \alpha \beta}(t)$ 
 in time, which will be used in later discussions.

 Firstly, we discuss the value of $\lambda$ at an initial time $t_0$, 
 at which the environmental branches possess
 no correlation with the operator $H^{I\E}$, e.g., in a thermal state.
 Clearly, such a state should possess negligible correlation with the operator $H^{I\E}$.
 In this case, as discussed previously, one has $\lambda =1$,
 which implies that $H^{I\E(4)}_{\phi, \alpha \beta}(t_0)$ should be negligibly
 small for a sufficiently large environment, scaling as $M_\Gamma^{-1/2}$.

 Secondly, we discuss the value of $\lambda$ after a finite time 
 period of evolution driven by the interaction Hamiltonian.
 Previously, it was argued that $\lambda$ should be equal to $1.5$,
 when the RDM undergoes a nonnegligible evolution beside phase changes.
 Now, with the environmental-branch evolution in Eq.(\ref{Phi(t)-exp}),
 one may understand the dynamic origin of this value of $\lambda=1.5$.
 For this purpose, according to Eq.(\ref{HIE4-M-scale}), one needs to show that 
 $H^{I\E(4)}_{\phi, \alpha \beta}(t)$ should be finite and nonegligible in the limit of large $N$.
 Note that, due to smallness of the contributions from $l=1,2,3$,
 this requires nonnegligibility of $H^{I\E}_{\phi, \alpha \beta}(t)$.

 To achieve the above discussed goal,  let us expand $H^{I\E}_{\phi, \alpha \beta}(t')$
 of $t'=t + \delta t$ in terms of $\delta t$
 and consider those first-order terms that contain $H^{I\E}$. 
 Making use of Eqs.(\ref{Phi(t)-exp}) and (\ref{HIE-phiab}), one gets that
\begin{align}\notag
  H^{I\E}_{\phi, \alpha \beta}&(t')  = H^{I\E}_{\phi, \alpha \beta}(t)
   -i \delta t \sum_{\gamma } H^{IS}_{\alpha \gamma } 
  H^{I\E2}_{\phi, \gamma \beta}(t)
  \\ & -i \delta t \sum_{\gamma } H^{IS}_{\gamma \beta} H^{I\E2}_{\phi, \alpha \gamma }(t) + \cdots
  \label{HIE-phiab-delta-t}
\end{align}
 where 
\begin{align}
 & H^{I\E2}_{\phi, \alpha \gamma }(t) \equiv \la \phi^\E_\gamma (t) | (H^{I\E})^2  |\phi^\E_\alpha (t)\ra.
 \label{HIE2-pab}
\end{align}
 Generically, the quantity $H^{I\E2}_{\phi, \alpha \gamma }(t)$ is not negligibly small.
 To see this point, let us use $h^{I\E 2}(e)$ to indicate the diagonal function 
 for the operator $O= (H^{I\E})^2$  in the ETH ansatz of Eq.(\ref{ETH}),
 and use $h^{I\E 2}_{0}$ to indicate its value at the center of the 
 energy window $\Gamma$. 
 Like $H^{I\E}_{\phi, \alpha \beta}$ in Eq.(\ref{HIE-phiab=sum-l}), 
 this quantity can also be divided into four parts labelled by $l$, i.e., 
 $H^{I\E2}_{\phi, \alpha \gamma }(t) = \sum_{l=1}^4 H^{I\E2 (l)}_{\phi, \alpha \gamma }(t)$.
 Then, from Eq.(\ref{HIE2-pab}), one gets the following expression for
 its $l=1$ part,
\begin{align}
 & H^{I\E2 (1)}_{\phi, \alpha \gamma }(t) = h^{I\E 2}_{0} 
 \rho^S_{\alpha \gamma}(t).
 \label{HIE2-pab-l=1}
\end{align}
 Note that, although the averaged diagonal part of $H^{I\E}$ can be 
 absorbed due to Hamiltonian renormalization, 
 usually, that of $(H^{I\E})^2$ can not. 
 In addition, this part is usually not small;
 for example, in a spin chain with $H^{I\E}$ as a Pauli matrix (say, $\sigma_x$), the operator $(H^{I\E})^2$ is equal to the unit matrix
 and $h^{I\E 2}_0 =1$.

 From Eqs.(\ref{HIE-phiab-delta-t})-(\ref{HIE2-pab-l=1}), one sees that 
 after a finite time evolution beyond the initial time $t_0$,
 $H^{I\E}_{\phi, \alpha \beta}(t)$ and hence 
 $H^{I\E(4)}_{\phi, \alpha \beta}(t)$ may get a notable value.
 Then, from the scaling behavior of $H^{I\E(4)}_{\phi, \alpha \beta}(t)$ 
 in Eq.(\ref{HIE4-M-scale}), one sees that $\lambda =1.5$ for $t$ beyond
 an initial time period.
 \footnote{The following is a heuristic argument for the value of $\lambda =1.5$.
 Clearly, it is impossible for the $2M_\Gamma$ phases of $C^*_{\beta j}(t)$ and $C_{\alpha i}(t)$ to fully compensate those of $H^{I\E}_{ji}$, whose number
 scales as $M^2_\Gamma$.
 In a possibly ultimate compensation,
 phases of $\big(\sum_i C_{\alpha i}(t) H^{I\E}_{ji} \big)$
 may match those of $C^*_{\beta j}(t)$ for all $j$,
 such that the sum over $j$ scales as $M_\Gamma$.
 In this case, phases of $C_{\alpha i}(t)$ may hardly match those of $H^{I\E}_{ji}$ for a fixed $j$;
 and, if completely nonmatching, the sum over $i$ may scale as $M_\Gamma^{1/2}$, 
 which would imply $\lambda =1.5$. 
 }

 Finally, we discuss time correlation of $H^{I\E (4)}_{\phi, \alpha \beta}(t)$.
 As seen from previous discussions, for the purpose of studying dynamics of the RDM of the central system,
 it is sufficient to consider a subspace of the environmental Hilbert space, 
 which is spanned by the $d_S$ environmental branches.
 We use $\HH^\E_{\{\phi \}}$ to denote this subspace,
 namely, $\HH^\E_{\{\phi \}} = \bigoplus_\alpha  |\phi^\E_\alpha(t)\ra $.
 Clearly, due to the relationship between the RDM elements and the overlaps of the environmental branches,
 dynamics of the RDM is determined by relative motion of the branches 
 within the subspace $\HH^\E_{\{\phi \}}$.

 Motion of the subspace $\HH^\E_{\{\phi \}}$ is governed by Eq.(\ref{Phi(t)-exp}),
 particularly, driven by $\M = [H^S] +  H^\E [I] +  H^{I\E}[H^{IS}]$.
 Due to the chaotic nature of the environment, usually, 
 the subspace $\HH^\E_{\{\phi \}}$ moves  in a irregular way
 within the whole state space of the environment. 
 This suggests decay of the time correlation of $H^{I\E (4)}_{\phi, \alpha \beta}(t)$.
 \footnote{The RDM, as determined by relative motion of the 
 environmental branches, may  behave less irregularly than the branches. }

 To be specific, let us consider the following correlation function,
 indicated as $C_{H\phi}^{I\E (4)}$,
\begin{align}
  C_{H\phi}^{I\E (4)}(t,t') = \ov{H^{I\E (4)}_{\phi, \alpha \beta}(t)
  H^{I\E (4)}_{\phi, \alpha \beta}(t')}.
\end{align}
 Let us write the $\alpha$-th row of $[|\phi^\E(t)\ra \big]$ as 
\begin{align} 
  \big[|\phi^\E(t)\ra \big]_\alpha & 
 = P_\alpha e^{-i \M (t-t_0)}\big[ |\phi^\E(t_0)\ra \big]
\label{}
\end{align}
 where $P_{\alpha}$ is a $d_S$-dimensional and one-row matrix, 
 $[0,\ldots,0, 1, 0,\ldots, 0]$, which has only one nonzero element 
 $1$ lying in the $\alpha$-th column. 
 In a similar way, the $\beta$-th column of $[\la \phi^\E(t)| \big]$ is written as
\begin{align}   \big[\la \phi^\E(t)| \big]_\beta & 
 = \big[ \la \phi^\E(t_0)| \big] e^{i \M (t-t_0)} P^\dag_\beta.
 \label{}
\end{align}
 In this notation, one writes that
\begin{subequations}\label{HIE-phiab-tt'}
 \begin{align}  \label{HIE-phiab-2}
  H^{I\E (4)}_{\phi, \alpha \beta}(t) & = \sum_{i \ne j} \big[ \la \phi^\E(t)| \big] 
  P^\dag_\beta | j\ra H^{I\E}_{ji} \la i| P_\alpha \big[ |\phi^\E(t)\ra \big],
 \\ \notag
  H^{I\E (4)}_{\phi, \alpha \beta}(t') & = \sum_{i \ne j} \big[ \la \phi^\E(t)| \big] 
 e^{i \M \Delta t} P^\dag_\beta | j\ra H^{I\E}_{ji} 
 \\ & \times \la i| 
 P_\alpha e^{-i \M \Delta t}\big[ |\phi^\E(t)\ra \big].
 \label{HIE-phiab-3}
\end{align}
\end{subequations}
 where $\Delta t = t' -t$.

 From the above expressions, one sees that 
 the difference between $H^{I\E (4)}_{\phi, \alpha \beta}(t)$
 and $H^{I\E (4)}_{\phi, \alpha \beta}(t')$ lies in that the latter
 contains the evolution operator $e^{-i \M \Delta t}$ and 
 its Hermitian conjugate.
 Due to the chaotic motion of the environment driven by $H^\E$, 
 the overall difference between $e^{-i \M \Delta t}\big[ |\phi^\E(t)\ra \big]$
 and $\big[ |\phi^\E(t)\ra \big]$ increases fast with $\Delta t$. 
 Moreover, since $i \ne j$ in Eq.(\ref{HIE-phiab-tt'}), 
 effects of $e^{i \M \Delta t}$ can not be completely compensated
 by those of $e^{-i \M \Delta t}$.
 These properties suggest that it should be reasonable to 
 assume that, typically, the correlation $C_{H\phi}^{I\E (4)}$
 may decay fast with increasing $\Delta t$. 
 We use $\tau_{{\rm corr}}^{\phi (4)}$ to indicate the decay time,
 beyond which phases of $H^{I\E (4)}_{\phi, \alpha \beta}$ 
 become effectively uncorrelated.

\section{Basic framework for an approach to master equation}
\label{sect-basic-framework}

 In this section, we discuss a method that can be 
 employed for deriving a master equation of RDM.
 We first discuss the basic strategy (Sec.\ref{sect-strategy-ME}),
 then, discuss the main approximations to be used (Sec.\ref{sect-b-form-ME}).

\subsection{Basic strategy of deriving master equation}\label{sect-strategy-ME}

 Suppose one is interested in the time evolution of the RDM of the system $S$
 within a finite time period of $[0,T]$, where $T$ is not very long.
 In order to derive a master equation for this period, 
 we are to adopt the following strategy.

 Firstly, the time period $[0,T]$ is divided into a series of $M$ short time intervals, 
 which have a length indicated by $\tau$.
 The time period is separated by instants $t_m = m \tau$ of  $m=0,\ldots, M$, 
 with $T = M \tau$.

 Secondly, within each short interval $(t_m, t_{m+1})$, 
 evolution of the RDM is studied, following a procedure similar to that 
 given in Sec.\ref{sect-evolv-t0-t} with $t_0$ replaced by $t_m$
 and $t$ by $t_{m+1}$.
 Here, one should note a difference between the environmental
 branches at $t_m$ of $m>0$ and those at the initial time $t_0$.
 That is, as discussed in Sec.\ref{sect-evolv-t0-t}, the initial 
 environmental branches are assumed to possess 
 no correlation with the operator $H^{I\E}$ such that $\lambda =1$;
 meanwhile, due to the time evolution under the interaction Hamiltonian,
 the environmental branches possess such correlations
 at instants $t_m$ of $m>0$, implying $\lambda=1.5$ as discussed previously,
 particularly in Sec.\ref{sect-HIE4-lambda}. 
 As a consequence, contributions from terms of $l=4$ are not  
 necessarily small at $t_m$.

 Finally, summing up results obtained for all the short time intervals,
 one studies whether a master equation may be obtained
 under certain approximations.

 To be specific, within each short interval, say, the $m$-th, 
 since $k_{\rm tru} \le k_{\rm ETH}$ for a large environment 
 as discussed previously,
 similar to the first equality in Eq.(\ref{rho(t)-G-t^k-t0}),
 variation of the RDM is written as
\begin{align}\label{drho-G-t^k-tm}
 & \rho^S_{\alpha \beta}(t_{m+1})  - \rho^S_{\alpha \beta}(t_{m})
 \simeq \sum_{k=1}^{k_{\rm tru}} G^{(k)}_{\alpha \beta}(t_{m}) \tau^{k},
\end{align}
 where $G^{(k)}_{\alpha \beta}(t_{m})$ has the same expression
 as the rhs of Eq.(\ref{Gk-sum-xYphi}) but with $t_0$ replaced by $t_m$.
 Further, with the division of $G^{(k)}_{\alpha \beta}(t_{m})$ 
 by the labels of $l$ and $\eta$ [cf.~Eq.(\ref{Omega-4})], one gets that
\begin{align}
 & \rho^S_{\alpha \beta}(T) - \rho^S_{\alpha \beta}(0)  \simeq
   \sum_{m=0}^{M-1} \sum_{k=1}^{k_{\rm tru}}
 \sum_{l=1}^4 \sum_\eta \tau^{k}  G^{(k,l)}_{\alpha \beta, \eta}(t_{m}).
\label{rho-T-sum-G}
\end{align}

Let us rewrite the above expression in the following form,
\begin{align}
 & \rho^S_{\alpha \beta}(T) \simeq \rho^S_{\alpha \beta}(0)
  + \sum_{m=0}^{M-1} \tau \LL_{\alpha \beta}(t_{m}),
\label{rho-T-LL-ab}
\end{align}
 where
\begin{subequations}\label{LL-ab-k}
\begin{align}
 & \LL_{\alpha \beta}(t_{m})  = \sum_{k=1}^{k_{\rm tru}} \LL^{(k)}_{\alpha \beta}(t_{m}),
\label{LL-ab}
 \\ &  \LL^{(k)}_{\alpha \beta}(t_{m}) =  
   {\sum_{l,\eta }}  \tau^{k-1}  G^{(k,l)}_{\alpha \beta, \eta}(t_{m}).
\label{LL-k}
\end{align}  
\end{subequations}
 In the operator form, Eq.(\ref{rho-T-LL-ab}) is written as
\begin{align}
 & \rho^S(T) - \rho^S(0)  \simeq
  \sum_{m=0}^{M-1} \tau \LL(t_m),
\label{rho-T-sum-LL}
\end{align}
 where 
\begin{align}\label{<a|LL|b>}
 \LL (t_m) = \sum_{\alpha \beta} |\alpha\ra \LL_{\alpha \beta}(t_m) \la \beta |.
\end{align}

 For $T$ not very large, due to smallness of $\tau$, 
 $(\rho^S(T) - \rho^S(0))$ can be approximately computed 
 by solving the following equation, 
\begin{align}\label{ME-formal}
 \frac{d\rho^S(t)}{dt} = \LL(t),
\end{align}
 where $\LL (t)$ is obtained by a continuous 
 extension of $\LL (t_m)$ in Eq.(\ref{<a|LL|b>}).
 One crucial point is to study whether $\LL(t)$ may act as 
 a superoperator on the RDM, written as $\LL(\rho^S(t))$.
 In a positive answer, Eq.(\ref{ME-formal}) would become a master equation.

\subsection{Main approximations to be used}
\label{sect-b-form-ME}

 In this section, we discuss main approximations that are to be used 
 when applying the above discussed strategy.
 This is to be done, based on environmental properties induced by its chaotic dynamics,
 particularly, the ETH ansatz.

 As seen in Eq.(\ref{LL-ab-k}), the main quantities concerned are
 $G^{(k,l)}_{\alpha \beta, \eta}(t_{m})$.
 As discussed previously, the ETH ansatz implies that these quantities
 show quite different behaviors at different values of $l$. 
 Note that the $l$-dependence of $G^{(k,l)}_{\alpha \beta, \eta}(t_m)$
 lies in the quantity of $Y^{(l)}_{\eta \phi, \gamma \gamma'}(t_m)$
 [cf.~Eq.(\ref{Gketa-l-sum-xYphi})],
 which is the $l$-th part of
 $Y_{\eta \phi, \gamma \gamma'}(t_m)=
 \la \phi_{\gamma'}^\E(t_m)|  Y_{\eta} |\phi_{\gamma}^\E(t_m)\ra$.
 Below, we discuss contributions from $l=1,2,3,4$ separately.

 Firstly, for $l=1$, 
 as discussed previously, $Y^{(1)}_{\eta \phi, \gamma \gamma'}$
 is related to a RDM element [Eq.(\ref{Yketa-phiab-(1)})].
 In fact, according to Eq.(\ref{Gk-l1-sum}), 
 $G^{(k,1)}_{\alpha \beta, \eta}(t)$ is a linear function of the RDM. 
 Hence, the contribution of this part takes a form, which is expected for 
 the rhs of Eq.(\ref{ME-formal}) to be of a master-equation form.

 Secondly, we discuss terms of $l=2$.
 As seen in Sec.\ref{sect-evolv-t0-t}, particularly Eq.(\ref{Yketa(2)-CCh'}), 
 these terms are proportional to the slopes $y'_{\eta 0}$ 
 of $Y_{\eta \phi, \gamma \gamma'}$, taken at the center of 
 the energy shell $\Gamma$.
 As discussed previously (see Secs.\ref{HIE-under-ETH}
 and \ref{sect-evolv-t0-t}),
 although the ETH ansatz conjectures smallness of these slopes,
 generically, it does not give a quantitative prediction which may enable 
 one to make a judgement on whether the smallness could be sufficient
 for neglecting the contribution of terms of $l=2$.

 For the above reason, as the first main approximation in addition to 
 the ETH ansatz, we introduce the following assumption, which is to 
 be called \textit{approximation of small slope}.
\begin{itemize}
  \item 
 Approximation of small slope:  
 For all those operators $Y_\eta$ that appear at $k \le k_{\rm tru}(t)$,
 the slopes of $y_{\eta}(e)$ are sufficiently small 
 in the energy region concerned, 
 such that contributions from $G^{(k,l)}_{\alpha \beta,\eta }(t_{m})$
 of $l=2$ are negligible.
\end{itemize}
 Specifically, the slopes are $y'_{\eta 0}$.

 Thirdly, for $l=3$, for the same reason as that discussed 
 below Eq.(\ref{HIE-phiab-(3)}), for a large environment, 
 all $G^{(k,l)}_{\alpha \beta, \eta}(t_{m})$ of $l=3$ are negligible, too.

 Fourthly, we discuss the case of $l=4$. 
 As discussed previously, with $\lambda =1.5$ (except at the initial time), 
 the quantities 
 $Y^{(k,4)}_{\eta \phi, \gamma \gamma'}$ are not necessarily small. 
 One key observation is that arguments given in Sec.\ref{sect-HIE4-lambda}
 for decay of phase correlation of $H^{I\E (4)}_{\phi, \alpha \beta}(t)$ in time
 is also applicable to $Y^{(k,4)}_{\eta \phi, \gamma \gamma'}$
 for other nontrivial operators $Y_\eta$,
 still due to chaotic dynamics of the environment.
 For brevity, we still use $\tau_{{\rm corr}}^{\phi (4)}$ to indicate 
 the maximum correlation time for all these operators;
 that is, the phase correlation of times $t$ and $t'$ 
 may be effectively neglected for $|t-t'| > \tau_{{\rm corr}}^{\phi (4)}$.

 Let us consider the final contribution
 of $G^{(k,4)}_{\alpha \beta,\eta}(t_{m})$ for the whole time period $T$.
\footnote{Note that there is no need to consider the case of 
$Y_\eta = I^\E$, 
 for which $f(e^0,\omega)=0$ and hence $G^{(k,4)}_{\alpha \beta, \eta}=0$.}
 This final contribution is written as
\begin{align}
 & \left| \sum_{m=0}^{M-1} G^{(k,4)}_{\alpha \beta, \eta}(t_{m}) \right|
 \simeq \sigma M^\xi ,
\label{rho-T-sum-G-k4-eta}
\end{align}
 where $\sigma^2$ represents the variance of $G^{(k,4)}_{\alpha \beta, \eta}(t_m)$
 and $\xi$ is some positive parameter, satisfying $\xi \le 1$.
 The value of $\xi$ depends on dynamics of the environment $\E$, 
 as well as on the value of $\tau$.
 For $\tau \ge \tau_{{\rm corr}}^{\phi (4)}$ under a quantum chaotic 
 environment, one has $\xi \simeq 1/2$.

 Let us compare contributions from $l=1$ and $l=4$.
 For $l=1$, at least for some pair of $(k,\eta)$, 
 say, for that with $Y_\eta = (H^{I\E})^2$,
 $G^{(k,1)}_{\alpha \beta, \eta}(t_m)$ should have an average value notably different than zero (at least, for $T$ not very long).
 This implies the following behavior of the sum of these terms over $m$, i.e., 
\begin{align}
 & \sum_{m=0}^{M-1} G^{(k,1)}_{\alpha \beta, \eta}(t_{m}) \simeq \kappa M, 
\label{rho-T-sum-G-k1-eta}
\end{align}
 where $\kappa$ indicates the average of $G^{(k,1)}_{\alpha \beta, \eta}$.
 From Eq.(\ref{rho-T-sum-G-k4-eta}) and Eq.(\ref{rho-T-sum-G-k1-eta})
 one sees that, for a sufficiently large value of $M$,
 it should be possible for the final contribution 
 from terms of $l=4$ to be much smaller than that for $l=1$.

 Based on the above discussions, we make the following approximation,
 which is to be called the \textit{approximation of branch-correlation decay}.
\begin{itemize}
  \item Approximation of branch-correlation decay.
 The time interval width $\tau$ is sufficiently large compared with 
 $\tau_{{\rm corr}}^{\phi (4)}$ and the value of $M$ is sufficiently large,
 such that the final contribution of the terms of $l=4$ 
 is much smaller than that for $l=1$.
\end{itemize}
 Thus, effectively (in an average sense) 
 the terms of $l=4$ can be neglected, when computing the final RDM.
 The exact validity condition for the approximation of branch-correlation decay
 should be case-dependent.
 \footnote{One example for this condition will be discussed later
 in Sec.\ref{sect-compare-l=1-4}.}

 To summarize, under the ETH ansatz, the approximation of small slope,
 and the approximation of branch-correlation decay, 
 in an effective sense, one may consider contributions 
 only from terms of $l=1$ in the computation of the time variation
 of the averaged RDM, indicated as $\ov \rho^S$.
 Thus, Eq.(\ref{rho(t)-G-t^k-t0}) 
 remains valid in the average sense, with $(t,t_0)$ replaced $(t_{m+1},t_m)$. 
 This gives the following expression for the average of 
 $\LL_{\alpha \beta}$, indicated as $\ov \LL_{\alpha \beta}$, namely, 
\begin{align} 
  & \ov\LL_{\alpha \beta}(t_{m}) \simeq \sum_{k=1}^{k_{\rm tru}} \sum_{\eta,\gamma,\gamma'} y_{\eta 0}
 x_{\alpha \beta, \eta}^{(k)  \gamma \gamma'}
 \ov \rho^S_{\gamma \gamma'}(t_m) \tau^{k-1}.
 \label{rho(t)-G-t^k-tm-2}
\end{align}
 As discussed previously, $x_{\alpha \beta, \eta}^{(k)\gamma \gamma'}$
 are initial-state independent and hence independent of the RDM.
 Meanwhile, the value of $y_{\eta 0}$ depends only on the 
 environmental energy region involved, not on details of the RDM. 
 Hence, basically, the rhs of Eq.(\ref{rho(t)-G-t^k-tm-2}) is a linear function
 of $\ov\rho^S(t_m)$.

 Thus, although not getting a master equation exactly in the form of Eq.(\ref{ME-formal}), we obtain a master equation for the averaged RDM,
 which is useful for computing $\rho^S(T)$ from $\rho^S(0)$ in an effective sense.
 Its discrete form is written as 
\begin{align}
 & \ov\rho^S(t_{m+1}) - \ov\rho^S(t_m)  \simeq \tau \ov\LL(\ov\rho^S(t_m)),
\label{rho-T-sum-LL-tm}
\end{align}
 and its continuous form as 
\begin{align}\label{ME-formal-ov}
 \frac{d \ov\rho^S(t)}{dt} = \ov\LL(\ov\rho^S(t)),
\end{align}
 whose rhs is determined by the rhs of Eq.(\ref{rho(t)-G-t^k-tm-2}). 
 In what follows, we mainly discuss $\ov \rho^S$ and $\ov \LL$ and,
 for brevity, the overlines above $\LL$ and $\rho^S$ are usually omitted
 when there is no risk of causing confusion.

 Remarks about the value of $\tau$:
 (i) The main reason of requiring smallness of $\tau$ is to derive 
 the continuous form of master equation in Eq.(\ref{ME-formal-ov}). 
 (ii) The approximation of branch-correlation decay sets a lower bound to 
 the value of $\tau$ and, 
 when $\tau$ is not very small, it is unnecessary for Eq.(\ref{ME-formal-ov}) to work well.
 However, the discrete master equation in Eq.(\ref{rho-T-sum-LL-tm}) does not 
 require such smallness of $\tau$. 
 Moreover, one notes that the value of $k_{\rm tru}$ on the rhs
 of Eq.(\ref{rho(t)-G-t^k-tm-2}) increases with the increase of $\tau$.

 It may be useful to mention another expression
 for $\LL_{\alpha \beta}(t_{m})$.
 That is, the labels $\gamma$ and $\gamma'$ on the rhs of 
 Eq.(\ref{rho(t)-G-t^k-tm-2}) can be suppressed
 by writing in the $\alpha$-matrix notation, i.e., 
\begin{align} 
  & \LL_{\alpha \beta}(t_{m}) \simeq \sum_{k=1}^{k_{\rm tru}} \sum_{\eta} 
  y_{\eta 0} \tau^{k-1}
 \tr \left( \big[x_{\alpha \beta, \eta}^{(k) }\big]^\dag \rho^S(t_m)  \right).
 \label{LL-t^k-tm-xrho}
\end{align}
 Furthermore, making use of Eq.(\ref{u*u-XX-phi}), 
 it is possible to rewrite the rhs of Eq.(\ref{LL-t^k-tm-xrho}) in a form, 
 in which the RDM does not appear in a trace;
 in this case what appear are terms like 
 $ \left(   [u^{k-n,r}] \rho^S(t_m) [u^{n,r' }]^\dag \right)_{\alpha \beta}$.

 Finally, we discuss again the initial condition.
 The above arguments for negligibility of contributions from $l=4$
 were based on loss of phase correlation of $H^{I\E (4)}_{\phi, \alpha \beta}$ 
 for $|t-t'|$ beyond $\tau_{{\rm corr}}^{\phi (4)}$.
 One should note that this loss is in fact an averaged effect
 and does not necessarily happen quickly for all initial conditions,
 though the chaotic dynamics of the environment may finally destroy 
 initial correlations under a generic type of interaction that is not too weak.
 It is for this reason that we consider an initial condition, in which  the initial
 environmental branches possess no correlation with the interaction Hamiltonian.

\section{Master equation with $k_{\rm tru}=2$}\label{sect-ME-ktru=2}

 In this section, as an application of the framework discussed above,
 we derive a master equation 
 in the simplest nontrivial case, namely for $k_{\rm tru} = 2$.
 Specifically, expressions of the $G$-functions needed for the derivation
 are given in Sec.\ref{sect-G-k=1}
 and the master equation is derived in Sec.\ref{sect-LL-tau}.
 An explicit condition for neglecting those contributions of $l=4$ 
 in this specific case is discussed in Sec.\ref{sect-compare-l=1-4}
 and an impact of this condition on the value of $\tau$ is discussed in Sec.\ref{sect-tau}.
 In Sec.\ref{sect-dephasing}, as an application,
 decoherence induced by nondissipative interactions (pure dephasing), 
 for which $H^{IS}_{\alpha  \beta } =0$ for all $\alpha \ne \beta$,
 is discussed and compared with prediction of the random matrix theory.

\subsection{Expressions of $G^{(k)}_{\alpha \beta, \eta}(t_m)$ of $k=1,2$}\label{sect-G-k=1}

 In this section, we discuss expressions of $G^{(k)}_{\alpha \beta, \eta}(t_m)$
 of $k=1,2$. 
 For this purpose, one needs to compute the related operators $Y_\eta$. 
 It turns out that only four such operators of $\eta =1,2,3,4$ are to appear,
 which are listed below ahead of time, 
\begin{align}\label{Yk-01}
 Y_{\eta} = \left\{   \begin{array}{ll}  I^\E, & \quad \text{for $\eta=1$ ($k=1,2$)};
  \\ H^{I\E}, & \quad \text{for $\eta=2$ ($k=1,2$)};
  \\   H^{I\E,\E} , & \quad \text{for $\eta=3$ ($k=2$)};
  \\ (H^{I\E})^2, & \quad \text{for $\eta=4$ ($k=2$)};
                 \end{array}    \right.
\end{align}
 where
\begin{align}\label{HIE,E}
  H^{I\E,\E} \equiv [H^{I\E}, H^\E ].
\end{align}

 For $k=1$, substituting Eq.(\ref{M-matrix}) into Eq.(\ref{Gkt0=sum-phi}) with $t_0$ replaced by $t_m$,
 it is direct to get $G^{(1)}_{\alpha \beta}(t_m)$ [see Eq.(\ref{G1-0-app}) in Appendix \ref{app-derive-Gk}],
\begin{align}
  G^{(1)}_{\alpha \beta}(t_m)  = \sum_{\eta=1}^2  G^{(1)}_{\alpha \beta, \eta}(t_m),
\label{G1-0}
\end{align}
 where 
\begin{subequations} \label{G1-0-4}
\begin{align}\label{G1-eta0}
 & G^{(1)}_{\alpha \beta 1}  (t_m)  =  i (e^S_{\beta} - e^S_{\alpha}) \rho^S_{\alpha \beta}(t_m),
  \\  &  G^{(1)}_{\alpha \beta 2}  (t_m) =
  i \sum_{\gamma } \Big( H^{IS}_{\gamma  \beta } H^{I\E}_{\phi, \alpha \gamma }(t_m)
   - H^{IS}_{\alpha \gamma  } H^{I\E}_{\phi, \gamma \beta}(t_m) \Big).
\label{G1-eta1}
\end{align}
\end{subequations}

 For $k=2$, one finds that (Appendix \ref{app-derive-Gk}),
\begin{align}\label{}\notag
  G^{(2)}_{\alpha \beta}(t_m) & =  \sum_{\gamma, \gamma'}
   \la \phi_\gamma^\E(t_m)|  \M_{ \gamma\beta}  \M_{\alpha \gamma'}  |\phi_{\gamma'}^\E(t_m)\ra
  \\ & \notag  - \frac 12 \sum_{\gamma} \la \phi_\gamma^\E(t_m)| \M^2_{ \gamma \beta} |\phi_\alpha^\E(t_m)\ra
  \\ & - \frac 12 \sum_{\gamma} \la \phi_\beta^\E(t_m)|  \M^2_{\alpha \gamma} |\phi_\gamma^\E(t_m)\ra .
  \label{G2=sum-M}
 \end{align}
 Inserting the $\alpha$-matrix elements
 of $ \M_{\alpha \gamma} = (e^S_{\alpha} +  H^\E) \delta_{\alpha \gamma}  +  H^{I\E} H^{IS}_{\alpha \gamma}$ [cf.~Eq.(\ref{M-matrix})]
 into Eq.(\ref{G2=sum-M}), one gets that 
 [see Eqs.(\ref{app-G2-l=0}) and (\ref{G2l-1,4})]
\begin{align}
   G^{(2)}_{\alpha \beta}(t_m) = \sum_{\eta =1}^4   G^{(2)}_{\alpha \beta, \eta}(t_m),
  \label{G21-Q-2HIS}
\end{align}
 where
\begin{subequations} \label{G21-Q-2HIS-4}
\begin{align}
  &   \label{G2-eta1}  G^{(2)}_{\alpha \beta, 1}(t_m)
   =   - \frac 12  (e^S_\beta - e^S_\alpha)^2  \rho^S_{\alpha \beta}(t_m),
 \\ &   \label{G2-eta2}  G^{(2)}_{\alpha \beta, 2}(t_m) =  Q,
 \\ &   \label{G2-eta3}  G^{(2)}_{\alpha \beta, 3}(t_m) =
       \frac 12  H^{IS}_{\gamma \beta} \, H^{I\E,\E}_{\phi, \alpha \gamma }(t_m)
    - \frac 12  H^{IS}_{\alpha \gamma}  H^{I\E,\E}_{\phi, \gamma \beta}(t_m),
 \\ & \notag  G^{(2)}_{\alpha \beta, 4}(t_m) =
  H^{IS}_{\gamma\beta}  H^{IS}_{\alpha \gamma'} H^{I\E2}_{\phi, \gamma' \gamma }(t_m)
 \\ &  \label{G2-eta4}   -\frac 12  (H^{IS})^2_{\gamma \beta} H^{I\E2}_{\phi, \alpha \gamma }(t_m)
  -\frac 12 (H^{IS})^2_{\alpha \gamma} H^{I\E2}_{\phi, \gamma \beta }(t_m).
\end{align}
\end{subequations}
 Here,
\begin{align}\notag
  Q = & H^{IS}_{\alpha \gamma} (e^S_\beta -\frac 12   (e^S_\gamma + e^S_\alpha ) ) 
 H^{I\E}_{\phi, \gamma \beta}(t_m)
  \\ &+ H^{IS}_{\gamma\beta}(e^S_\alpha -\frac 12  (e^S_\gamma + e^S_\beta ) ) 
  H^{I\E}_{\phi, \alpha\gamma}(t_m),
\end{align}
 $H^{I\E,\E}_{\phi, \alpha \gamma }(t) \equiv \la \phi^\E_\gamma (t)| H^{I\E,\E}  |\phi^\E_\alpha (t)\ra$,
 and $H^{I\E2}_{\phi, \alpha \gamma }(t)$ was defined in Eq.(\ref{HIE2-pab}).

\subsection{Derivation of master equation}\label{sect-LL-tau}

 In this section, making use of the $G$-functions gotten above, 
 we derive a master equation by computing $\LL_{\alpha \beta}(t_{m})$
 in Eq.(\ref{LL-ab-k}) with terms of $l=1$ only.

 We first discuss the case of $k=1$.
 For $\eta =1$ with $Y_1$ as the identity operator $I^\E$, from Eq.(\ref{G1-eta0}) one finds that
\begin{align}\label{}
 &  G^{(1,1)}_{\alpha \beta 1} = G^{(1)}_{\alpha \beta 1}(t_m)
  = i   \big[ \rho^S(t_m), H^{S} \big]_{\alpha \beta}.
\label{G1-l=0}
\end{align}
 For $\eta =2$ with $Y_2 = H^{I\E}$,
 from Eq.(\ref{G1-eta1}) one finds that 
\begin{align}
 G^{(1,1)}_{\alpha \beta 2}(t_m) =0.  
\end{align} 
 due to Hamiltonian renormalization.

 Next, we discuss the case of $k=2$.
 For $\eta=1$, from Eq.(\ref{G2-eta1}) one gets that
\begin{align}
 & G^{(2,1)}_{\alpha \beta 1}(t_m) 
   =   - \frac 12  (e^S_\beta - e^S_\alpha)^2  \rho^S_{\alpha \beta}(t_m).
 \label{Gab-21-1}
\end{align}
 For $\eta=2$, similar to the above case of $k=1$, one also has
\begin{align}
 & G^{(2,1)}_{\alpha \beta 2}(t_m) =0.
 \label{Gab-21-2=0}
\end{align}
 \\ For $\eta=3$ with $Y_\eta = [H^{I\E}, H^\E ]$,
 substituting the ETH ansatz Eq.(\ref{ETH}) with $O=H^{I\E}$ into 
 $\la i|[H^{I\E}, H^\E ]|j\ra = (e^\E_j - e^\E_i) \la i| H^{I\E} |j\ra $,
 with $f(e^0,\omega)$ standing for the offdiagonal function related to 
 $H^{I\E}$, one finds that
\begin{align}\label{}
 H^{I\E,\E}_{ij} = \frac{e^\E_j - e^\E_i}{\sqrt{\rho^\E_{\rm dos}(e^0)}} f(e^0,\omega) r_{ij}.
\end{align}
 It is seen that all diagonal elements of $H^{I\E,\E}_{ii}$ are equal to zero.
 As a result, one gets that
\begin{align}
 & G^{(2,1)}_{\alpha \beta 3}(t_m) = 0. 
 \label{Gab-21-3}
\end{align}
 For  $\eta=4$, making use of Eq.(\ref{HIE2-pab-l=1}),
 it is direct to compute $G^{(2,1)}_{\alpha \beta 4}(t_m)$ 
 from Eq.(\ref{G2-eta4}), getting that
\begin{align}\notag
  & G^{(2,1)}_{\alpha \beta 4}(t_m)    \simeq h^{I\E 2}_{0}
    H^{IS}_{\alpha \gamma'} \rho^{S}_{\gamma' \gamma}(t_m) H^{IS}_{\gamma\beta}
  \\ \notag & -\frac 12 h^{I\E 2}_{0} \rho^{S}_{\alpha \gamma}(t_m) (H^{IS})^2_{\gamma \beta}
   -\frac 12 h^{I\E 2}_{0} (H^{IS})^2_{\alpha \gamma} \rho^{S}_{\gamma \beta}(t_m)
 \\ & = h^{I\E 2}_{0} \Big( H^{IS} \rho^{S}(t_m) H^{IS}   -\frac 12 \{ \rho^{S}(t_m), (H^{IS})^2 \}  \Big)_{\alpha \beta}.
 \label{Gab-21-4}
\end{align}

 From the rhs of Eqs.(\ref{G1-l=0}) and (\ref{Gab-21-1}),
 one observes that a master equation, which generates 
 $G^{(1,1)}_{\alpha \beta 1}(t_{m})  \tau$  as the first-order term, 
 would generate $G^{(2,1)}_{\alpha \beta 1}(t_{m}) \tau^2$  as the second-order term.
 This implies that the term $G^{(2,1)}_{\alpha \beta 1}(t_{m})$
 may be neglected for the purpose of deriving a master equation.

 Finally, only two $G$-terms are left when computing
 $\LL_{\alpha \beta}(t_{m})$ by Eq.(\ref{LL-ab}),  namely,
 $G^{(1,1)}_{\alpha \beta 1}(t_m)$ and $\tau G^{(2,1)}_{\alpha \beta 4}(t_m)$.
 This gives the following superoperator for the master equation 
 in Eq.(\ref{ME-formal-ov}),
\begin{align}\label{}
  \LL = \LL^{(1)} + \LL^{(2)},
\label{LLk12}
\end{align}
 where
\begin{subequations}
\begin{align}\label{}
  & \LL^{(1)}(\rho^S) = i   \big[ \rho^S, H^{S} \big],
\label{LLk-l=0}
\\  & \LL^{(2)}(\rho^S)
 =  \tau h^{I\E 2}_{ 0} \left( H^{IS} \rho^S H^{IS}   -\frac 12  \{\rho^S, (H^{IS})^2 \} \right).
\label{LL-k=2}
\end{align}
\end{subequations}
 It is easy to see that the obtained master equation is of the Lindblad form. 

\subsection{Comparison between contributions from $l=1$ and $l=4$}
\label{sect-compare-l=1-4}

 In this section, we give a comparison between the contribution
 from $l=1$ and that from $l=4$, the latter of which has been neglected in the 
 above derivation of master equation. 
 In particular, a condition is to be derived under 
 which the latter is much smaller than the former.

 For $l=1$, the first-order term ($k=1$) generates a phase,
 as seen from Eq.(\ref{G1-l=0}), and hence has a quite small average value
 $\kappa$ for the rhs of Eq.(\ref{rho-T-sum-G-k1-eta}).
 For this reason, what needs consideration is the second-order 
 term ($k=2$), namely, $\tau^2 G^{(2,1)}_{\alpha \beta 4}(t_{m})$.
 According to Eq.(\ref{rho-T-sum-G-k1-eta}), 
 the final contribution from $\tau G^{(2,1)}_{\alpha \beta 4}(t_m)$ 
 is written as
\begin{align}
 & \tau^2  \sum_{m=0}^{M-1} G^{(2,1)}_{\alpha \beta 4}(t_{m}) = 
 \kappa \tau^2 M,
\label{rho-T-sum-G-k2}
\end{align}
 with $\kappa$ as the average value of $G^{(2,1)}_{\alpha \beta 4}$.

 On the other hand, for $l=4$, it is the variance $\sigma$
 that appears on the rhs of Eq.(\ref{rho-T-sum-G-k4-eta}).
 Here, it is usually sufficient to consider the first-order term ($k=1$).
 According to Eq.(\ref{Yk-01}), at $k=1$, $\eta$ has only two values,
 $\eta = 1,2$.
 Since the identity operator of $Y^{(1)}_{1} = I^\E$ has zero offdiagonal elements, what needs consideration is the case of $\eta=2$, more exactly,
 $\tau G^{(1,4)}_{\alpha \beta 2}(t_m)$.
 From Eq.(\ref{G1-eta1}), one gets that
\begin{align}
 &  G^{(1,4)}_{\alpha \beta 2}(t_m) 
 = i   \sum_{\gamma } \Big( H^{IS}_{\gamma  \beta }
    H_{\phi, \alpha \gamma }^{I\E (4)}(t_m)  - H^{IS}_{\alpha \gamma  }
    H_{\phi, \gamma \beta}^{I\E (4)}(t_m)  \Big),
\label{Gk1-l=4}
\end{align}
where $H_{\phi, \alpha \gamma }^{I\E (4)}(t_m)$ is given by Eq.(\ref{HIE-phiab-(4)}) with $t=t_m$.
 According to Eq.(\ref{rho-T-sum-G-k4-eta}),
 its final contribution  is written as
\begin{align}
 & \left| \sum_{m=0}^{M-1} \tau G^{(1,4)}_{\alpha \beta 2}(t_{m}) \right| \simeq \tau \sigma M^\xi ,
\label{rho-T-sum-G-k1}
\end{align}
 where $\sigma^2$ represents the variance of $G^{(1,4)}_{\alpha \beta 2}(t_m)$.

 Note that both $\sigma$ and $\kappa$ have negligible scaling
 dependence on the density of states, at least in the limit of large $N$.
 Indeed, from Eq.(\ref{Gk1-l=4}), one sees that
 the scaling behavior of $\sigma$ with respect to $\rho_{\rm dos}^\E$ is mainly determined by that
 of $H_{\phi, \alpha \beta }^{I\E (4)}$,
 which is $ H^{I\E (4)}_{\phi, \alpha \beta}  \sim (\rho_{\rm dos}^\E)^{0} $
  for $\lambda = 1.5$ in Eq.(\ref{HIE4-M-scale}).
 Meanwhile, the scaling behavior of $\kappa$ is
 mainly determined by that of $h^{I\E 2}(e^\E)$ [Eq.(\ref{Gab-21-4})];
 it behaves as $(\rho_{\rm dos}^\E)^{0} $,
 which is seen clearly from Eq.(\ref{HIE2ii-avg}) to be derived later
 due to the summation over $j$.

 Comparing Eq.(\ref{rho-T-sum-G-k1}) and Eq.(\ref{rho-T-sum-G-k2}), one sees that,
 if $\tau$ and $M$ satisfies the following relation,
\begin{align}
  \tau  \gg \frac{\sigma }{\kappa M^{1-\xi }}
 \Longleftrightarrow  \tau^\xi  \gg \frac{\sigma }{\kappa T^{1-\xi }},
\label{sigma1-g0}
\end{align}
 then,
\begin{align}
\label{G>G}
  \tau \left|\sum_{m=0}^{M-1} G^{(2,1)}_{\alpha \beta 4}(t_m) \right|
 \gg \left| \sum_{m=0}^{M-1} G^{(1,4)}_{\alpha \beta2}(t_m) \right|.
\end{align}
 For $\xi <1$, Eq.(\ref{sigma1-g0}) and hence Eq.(\ref{G>G}) 
 can be satisfied for a sufficiently large value of $M$,
 as a consequence, the final contribution from $G^{(1,4)}_{\alpha \beta 2}$ can be neglected.
 In particular, as discussed previously, 
 $\xi \simeq 1/2$ for $\tau \ge \tau_{{\rm corr}}^{\phi (4)}$.
 Hence, if $\tau_{{\rm corr}}^{\phi (4)}$ is sufficiently small,
 the final contribution from $G^{(1,4)}_{\alpha \beta 2}$ could be negligible
 with a small $\tau$.
 \footnote{If the value of $\tau$ needed is not sufficiently small, 
 one may need to consider $k_{\rm tru}$ larger than $2$.}

\subsection{A scale estimate to $\tau$}\label{sect-tau}

 In this section, we discuss a rough estimate to a scaling property
 of the time interval $\tau$, which is imposed by the condition
 in Eq.(\ref{sigma1-g0}).

 For this purpose,  one needs to give an estimate to the average value
 $\kappa$ of $G^{(2,1)}_{\alpha \beta 4}$,
 which  according to Eq.(\ref{Gab-21-4}) is related to 
 the diagonal function $h^{I\E 2}(e)$ for $(H^{I\E})^2$.
 Let us substitute the ETH ansatz Eq.(\ref{ETH}) 
 with $O=H^{I\E}$ into $(H^{I\E})^2_{ii} = \sum_{j} H^{I\E}_{ij} H^{I\E}_{ji} $,
 still using $f(e^0,\omega)$ to represent the offdiagonal function for $H^{I\E}$.
 Noting that the diagonal parts for $H^{I\E}$ are in fact negligible in this computation, one gets the following expression,
\begin{equation}\label{HIE2ii-avg}
 h^{I\E 2}(e^\E_i) \simeq  \sum_{j} \frac{1}{\rho^{\E}_{\rm dos}(e^0)} f^2(e^0,\omega),
\end{equation}
 where $ |r_{ii'}|^2$ has been replaced by $1$, since we are computing 
 the averaged diagonal elements.

 We recall the following feature of the function $f(e^0,\omega)$
 \cite{Rigol-AiP16,Deutch-RPP18,Dymarsky22BoundETH,pre25-ETHsc,WW-ETH-conf}.
 That is, it decays relatively slowly within a central region of 
 $\omega = e^\E_i - e^\E_j$ around $\omega=0$, with a width $w_f$;
 while, it decays exponentially beyond the central region.
 Hence, a rough estimate given by Eq.(\ref{HIE2ii-avg}) is
\begin{align}\label{hIE2-wff0}
 h^{I\E 2}(e^\E_i)  \simeq   w_f f^2_0,
\end{align}
 where $f_0$ indicates the average value of  $f(e^0,\omega)$ within the central region.

 Since $M \gg 1$, a scaling estimate to $\tau$ from Eq.(\ref{sigma1-g0}) is given by
 $\tau \sim \sigma / \kappa$.
 From Eqs.(\ref{Gab-21-4}) and (\ref{hIE2-wff0}),  one sees that $\kappa$ takes the following from
\begin{align}\label{}
 \kappa = w_f f^2_0 F_\kappa(H^{IS},\rho^S),
\end{align}
 where $F_\kappa (H^{IS},\rho^S)$ represents the contribution from $H^{IS}$ and $\rho^S$,
 whose dependence on $\rho_{\rm dos}^\E$ is negligible at least in a scaling sense.

 To get an estimate to $\sigma^2$,
 as the variance of $G^{(1,4)}_{\alpha \beta 2}(t_m)$ in Eq.(\ref{Gk1-l=4}), 
 we note that the variance of $H^{I\E}_{ji}$ should contain $f_0^2$ according to the ETH ansatz.
 Then, one may write
\begin{align}\label{}
 \sigma = f_0 F_\sigma (H^{IS},\rho^S),
\end{align}
 where $F_\sigma (H^{IS},\rho^S)$ represents the rest contribution,
 whose dependence on $\rho_{\rm dos}^\E$ is also negligible at least in a scaling 
 sense due to $\lambda = 1.5$.
 Finally, one gets that
\begin{align}\label{tau-estim}
 \tau \sim \frac{\sigma}{\kappa} \sim \frac{1}{w_f}.
\end{align}

\subsection{An application to decoherence }\label{sect-dephasing}

 In this section, we first show that the theory discussed in previous sections 
 indeed predicts decoherence 
 for a generic (small) system $S$ under generic nondissipative interactions 
 (Sec.\ref{sect-result-here}).
 Then, we give a comparison between the prediction here and
 what is known in the specific situation in which the
 RMT is applicable (Sec.\ref{sect-result-RMT}).

\subsubsection{Prediction of the theory here}\label{sect-result-here}

 For the sake of simplicity in discussion, we neglect variation 
 of the density of states within the energy window concerned.
 Making use of Eq.(\ref{ME-formal-ov}) with $\LL$ in Eq.(\ref{LLk12}) and noting Eq.(\ref{hIE2-wff0}),
 one gets the following master equation
 for offdiagonal elements $\rho^S_{\alpha \beta}$ with $\alpha \ne \beta$,
\begin{align}\label{}
    & \frac{d \rho^S_{\alpha \beta}}{d t}
     \simeq i (e^S_\beta - e^S_\alpha) \rho^S_{\alpha \beta}
     - \frac{g}{2} \Delta_{\alpha \beta}^2  \rho^S_{\alpha \beta},
\label{drho-dt-nondisp}
\end{align}
 where
\begin{align}
 & \Delta_{\alpha \beta} =  H_{\beta \beta}^{IS} - H_{\alpha \alpha}^{IS},
\label{Delta-ab}
 \\ & g=  w_f f^2_0 \tau.
 \label{g=g0-tau}
\end{align}

 It is easy to solve the above equation and get
 the following exponential decay for the offdiagonal RDM elements,
\begin{align}\label{}
    & |\rho^S_{\alpha \beta}(t)| =
   \exp \left(  - R_{\rm dec}  t \right) |\rho^S_{\alpha \beta}(t_0)|,
\label{rhot-pureD}
\end{align}
 where $R_{\rm dec}$ is the decoherence rate, 
\begin{align}\label{}
 R_{\rm dec} = \frac{g}{2} \Delta_{\alpha \beta}^2.
 \label{Rd1}
\end{align}
 One should note that the above prediction of decoherence, 
 though in consistency with the exponential decay predicted by pure dephasing, 
 is derived under a dynamic environment as a generic many-body 
 quantum chaotic system. 

\subsubsection{Comparison with result of random-matrix theory}\label{sect-result-RMT}

 Under nondissipative $S$-$\E$ interactions, offdiagonal elements of the RDM
 take the form of the so-called quantum Loschmidt echo (LE),
 which was first introduced as a measure for the sensitivity of quantum motion to small perturbation
 \cite{peres84}.
 In fact, under a nondissipative interaction,
 Eq.(\ref{dphi-dt}) for the evolution of environmental branches 
 reduces to a form of the Schr\"{o}dinger type
 and predicts the following unitary evolution,
\begin{align}\label{}
 |\phi_\alpha^\E(t)\ra = e^{-i H^{\rm eff}_\alpha (t-t_0)} |\phi_\alpha^\E(t_0) \ra,
\end{align}
 where  $H^{\rm eff}_\alpha$ is an effective Hamiltonian,
\begin{align}\label{}
    H^{\rm eff}_\alpha = e^S_\alpha + H^\E + H^I_{\alpha \alpha}. \label{Heff-alpha}
\end{align}
 Then, according to Eq.(\ref{rhoab-phiba}),
\begin{align}\label{rho-LE}
 \rho^S_{\alpha \beta}(t) = \la \phi_\alpha^\E(t_0) | e^{i H^{\rm eff}_\beta (t-t_0)}
 e^{-i H^{\rm eff}_\alpha (t-t_0)} |\phi_\alpha^\E(t_0) \ra.
\end{align}
 The quantity on the rhs of Eq.(\ref{rho-LE}) takes the form of LE, under a perturbation given by
\begin{align}\label{rho-LE-V}
 V = H^I_{\beta\beta} - H^I_{\alpha\alpha} = \Delta_{\alpha \beta} H^{I\E}.
\end{align}

 Decaying behaviors of the LE in quantum chaotic systems 
 have been studied in the field of quantum chaos by several methods,
 mainly by the semiclassical theory, a linear response theory, and the RMT
 \cite{jal2001,CT02,jac2001,cuc2002,prosen2002,STB03,GPSZ06,WCL04,WL05}.
 In particular, in the so-called Fermi-golden-rule regime of perturbation strength,
 different approaches give consistent results.
 For the purpose of comparing with the prediction in Eq.(\ref{rhot-pureD}),
 it is convenient to consider the approach of RMT,
 which gives the following result \cite{jac2001,cuc2002,GPSZ06},
\begin{align}\label{}
|f_{\beta \alpha }(t)| \sim \exp \left( - R^{\rm RMT}_{\rm dec} t \right),
 \label{ft-fgr}
\end{align}
 where
\begin{align}\label{}
 R^{\rm RMT}_{\rm dec} = \frac{\pi  \overline{ V_{nd}^2} }{\Delta } .
 \label{Rd}
\end{align}
 Here, $\Delta $ is the mean level spacing of the 
 effective Hamiltonian $H^{\rm eff}_{\alpha }$
 and $\overline{ V_{nd}^2}$ is an average for the offdiagonal elements of $V$
 on the eigenbasis of $H^{\rm eff}_{\alpha }$.
 For a random matrix of $H^{I\E}$, which clearly satisfies the ETH ansatz, with the constant offdiagonal function
 also denoted by $f_0$, direct computation for $V$ in Eq.(\ref{rho-LE-V}) shows that
 $\overline{ V_{nd}^2} = \Delta_{\alpha \beta}^2 f_0^2 / \rho_{\rm dos}^\E$.
 Then, noting that $\rho_{\rm dos}^\E = 1/ \Delta $, one gets that
\begin{align}\label{}
 R^{\rm RMT}_{\rm dec} =\pi  \Delta_{\alpha \beta}^2 f_0^2.
 \label{Rd1}
\end{align}

 As is known, prediction given by the RMT for observable elements 
 can be regarded as a special case dealt with by the ETH ansatz. 
 \footnote{Unlike the ETH ansatz, the RMT does not deal with any specific
 dynamic property of local observables. }
 Indeed, the decay in Eq.(\ref{ft-fgr}) becomes that in Eq.(\ref{rhot-pureD}),
 under an appropriate choice of $\tau$, denoted by $\tau_{\rm RMT}$.
 Note that $w_f$ for random matrices should be equal to
 their energy domain, denoted by $\Delta E$.
 Then, with $\hbar$ written explicitly, one finds that
\begin{align}\label{tau-RMT}
 \tau_{\rm RMT} = \frac{ 2\pi \hbar}{\Delta E }.
\end{align}
 Clearly, this is in consistency with the estimate in Eq.(\ref{tau-estim}).

\section{A refined framework}\label{sect-refinement}

 In this section, we discuss a refinement of the basic framework 
 proposed in Sec.\ref{sect-basic-framework}. 
 Firstly, in Sec.\ref{sect-shortcoming}, 
 we discuss a treatment employed in the basic framework,
 which was used for the sake of simplicity in discussion
 while may impose unnecessary restriction to the applicability 
 of the approximation of small slope.
 Then,  a method is discussed in Sec.\ref{sect-refine-l},
 which may improve the treatment and give rise to a refined framework. 
 Finally, some application of the refined framework is discussed in
 Sec.\ref{sect-refine-application}. 

\subsection{About a simplified treatment used in the basic framework}\label{sect-shortcoming}

 From discussions given previously, one sees that
 a basic strategy employed in the proposed framework is to divide
 each quantity $Y_{\eta \phi, \alpha \beta}$
 into four parts labelled by $l$ [Eq.(\ref{Yphi})], 
 based on predictions of the ETH ansatz.
 One key point lies in that the four parts behave quite differently.
 In particular, 
 a main purpose of introducing the approximation of small slope,
 which assumes sufficient smallness of the slopes $y'_{\eta 0}$,
 is to suppress the contribution from terms of $l=2$ such that 
 it can be neglected.

 However, one notes that, beside the slope $y'_{\eta 0}$,
 $Y^{(2)}_{\eta\phi, \alpha \beta}(t)$ in Eq.(\ref{Yketa(2)-CCh'})
 also contains the energy difference $(e^\E_i - e^\Gamma_{c})$. 
 We recall that one main reason of considering $e^\Gamma_{c}$
 (the center of the shell $\Gamma$) is that the value of $y_{\eta 0}$, 
 which is taken at $e^\Gamma_{c}$, is independent
 of the two labels $\alpha $ and $\beta $.
 More exactly, this independence allows movement of
 certain averaged effect of the system-environment interaction into
 the self-Hamiltonian of the system $S$ by taking a 
 Hamiltonian renormalization.

 As a cost, when the energies $e^\E_i$ of large components of 
 $C^*_{\beta i}(t)$ and $C_{\alpha i}(t)$ are not around $e^\Gamma_{c}$,
 $e^\Gamma_{c}$ is not the best choice
 for the purpose of suppressing the sum on the rhs of Eq.(\ref{Yketa(2)-CCh'})
 for $Y^{(2)}_{\eta\phi, \alpha \beta}(t)$.
 In fact, from Eq.(\ref{Yketa-phiab-(2)}), one sees that
 $Y^{(2)}_{\eta\phi, \alpha \beta}(t)=0$, if $y_{\eta 0} 
 = y_\eta(e^\Gamma_{c})$ is replaced by 
 $y_{\eta}(e_1)$ where $e_1$ is determined by the following relation, 
\begin{align}
 &  \sum_{i}  C^*_{\beta i}(t) C_{\alpha i}(t) y_{\eta}(e^\E_i)
  = y_{\eta}(e_1) \sum_{i } C^*_{\gamma' i}(t) C_{\gamma i}(t).
  \label{Y(k)eta-e1}
\end{align} 
 However, such a value of $e_1$ is usually not attractive, since it
 depends on the time $t$ in a complicated way and, 
 hence, $y_{\eta}(e_1)$ is usually not 
 a good parameter for a master equation.

 Therefore, one should balance between 
 the requirement of suppressing the contribution of $l=2$
 and that parameters used in a master equation 
 should not behave in a too complicated manner. 
 Clearly, the basic framework does not treat this balance in the best way. 
 As an improvement, we are to consider an $l$-label division,
 which is done with respect to centers of 
 the shells $\Gamma_{\alpha \beta} \equiv \Gamma_\alpha \cap \Gamma_\beta$,
 a topic to be discussed in detail in the next section.

\subsection{Refinement on the $l$-label division}\label{sect-refine-l}

 In order to distinguish between the $l$-label division to be discussed below and that used in the basic framework,
 we use $\ww l$ with a tilde to indicate the former
 which is done with respect to the center of $\Gamma_{\alpha \beta}$.
 Changes that are brought by $\ww l$ are also indicated by tilde,
 e.g., the superoperator to be derived is to be written as $\ww \LL$.

 We use $y_{\eta,\alpha \beta 0}$ to indicate the value of the ETH diagonal 
 function for $Y_\eta$ at the center of $\Gamma_{\alpha \beta}$, namely, 
\begin{align}
 y_{\eta,\alpha \beta 0} = y_\eta(e_{0}) \quad
 \text{with $e_{0}$ at the center of $\Gamma_{\alpha \beta}$}.
 \label{hIE2-ab0-ww}
\end{align}
 But, for the two operators $H^{I\E}$ and $H^{I\E 2}$, 
 to be in consistency with previous notations, we use $h^{I\E}_{\alpha \beta 0}$
 and $h^{I\E 2}_{\alpha \beta 0}$ standing for $y_{\eta,\alpha \beta 0}$,
 respectively.
 For $Y_{\eta \phi, \alpha \beta}$ in Eq.(\ref{Yphi}), 
 the new division is written as
\begin{align}
 Y_{\eta \phi, \alpha \beta} = \sum_{l=1}^4 
 Y^{(\ww l)}_{\eta \phi, \alpha \beta}.
 \label{hIE-ab0}
\end{align}
 where $\ww l$ indicates the following contributions:
\begin{subequations}\label{ww-l=1,4}
\begin{align}
 & \text{$\ww l=\ww 1$: from $y_{\eta,\alpha \beta 0}$ of the ETH diagonal function, }
 \\  & \text{$\ww l= \ww 2$: from deviation of  $y_{\eta}(e)$ from $y_{\eta,\alpha \beta 0}$, }
 \\  & \text{$\ww l =\ww 3$: from fluctuations of diagonal elements,}
 \\  & \text{$\ww l = \ww 4$: from ETH offdiagonal elements.}
\end{align}
\end{subequations}

 Let us first reconsider the simple case of $k_{\rm tru}=2$, 
 which was discussed within the basic framework in Sec.\ref{sect-ME-ktru=2}.
 Similar to Eq.(\ref{Yketa-phiab-(1)}),  one finds that
\begin{subequations}
\begin{align}
 & H^{I\E (\ww 1)}_{\phi, \alpha \beta }(t) = h^{I\E}_{\alpha \beta 0} 
 \rho_{\alpha \beta}^S(t) \equiv \ww\rho^{S,h1}_{\alpha \beta},
 \label{HIE1-g0-rho}
 \\ & H^{I\E2 (\ww 1)}_{\phi, \alpha \beta }(t) = h^{I\E 2}_{\alpha \beta 0} 
 \rho_{\alpha \beta}^S(t) \equiv \ww\rho^{S,h2}_{\alpha \beta}.
 \label{HIE2-g0-rho}
\end{align}  
\end{subequations}
 Here, we have introduced two notations of $\ww\rho^{S,h1}_{\alpha \beta}(t_m)$
 and $\ww\rho^{S,h2}_{\alpha \beta}(t_m)$,
 which are to be called \textit{rescaled RDMs}.

 Following procedures given previously when deriving
 Eq.(\ref{LLk12}), one gets a superoperator $\ww \LL$
 for a master equation, written as
 \begin{align}\label{}
 \ww \LL = \sum_{k=1}^{k_{\rm tru}} \ww \LL^{(k)}.
\label{LLk12-ww}
\end{align}
 Note that the Hamiltonian renormalization remains the same as before,
 with the system $S$'s renormalized Hamiltonian still taken as $H^S_{\rm rn}$
 in Eq.(\ref{OS-2}). 
 In the $\ww l$-label division, $H^{I\E (\ww 1)}_{\phi, \alpha \beta }(t)$
 can not be completely eliminated by the Hamiltonian renormalization.
 As a consequence, in addition to $\LL^{(1)}$, 
 $\ww \LL^{(1)}$ should contain a new term
 which comes from deviation of the renormalized 
 $h^{I\E}_{\alpha \beta 0}$ from $0$.
 More exactly, 
\begin{align}\label{}
  & \ww \LL^{(1)}(\rho^S) = i \big[ \rho^S, H^{S} \big]
  + i \big[ \ww\rho^{S,h1}, H^{IS} \big],
\label{LLk-l=0-ww}
\end{align}
 where $\ww\rho^{S,h1}$ is the rescaled RDM defined by Eq.(\ref{HIE1-g0-rho}).
 Note that values of $h^{I\E }_{\alpha \beta 0}$ are usually small due to the Hamiltonian renormalization.

 To compute $\ww \LL^{(2)}$, from Eq.(\ref{G2-eta4}), one finds that
\begin{align}\notag
  & G^{(2,1)}_{\alpha \beta 4}(t_m)  =
    H^{IS}_{\alpha \gamma'} \ww\rho^{S,h2}_{\gamma' \gamma}(t_m) H^{IS}_{\gamma\beta}
  \\ \notag & -\frac 12  \ww\rho^{S,h2}_{\alpha \gamma}(t_m) (H^{IS})^2_{\gamma \beta}
   -\frac 12  (H^{IS})^2_{\alpha \gamma} \ww\rho^{S,h2}_{\gamma \beta}(t_m)
 \\ & = \Big( H^{IS} \ww\rho^{S,h2}(t_m) H^{IS}   -\frac 12 \{ \ww\rho^{S,h2}(t_m), (H^{IS})^2 \}  \Big)_{\alpha \beta},
 \label{Gab-21-4-ww}
\end{align}
 where $\ww\rho^{S,h2}$ is the rescaled RDM defined by Eq.(\ref{HIE2-g0-rho}).
 Then, one gets that 
\begin{align}
 & \ww\LL^{(2)}(\rho^S)
 =  \tau H^{IS} \ww\rho^{S,h2} H^{IS}   -\frac{\tau}{2} \{ \ww\rho^{S,h2}, (H^{IS})^2 \}.
\label{LL-k=2-ww}
\end{align}

 Next, we discuss a generic value of $k_{\rm tru}$.
 In the $\ww l$-label division, similar to Eq.(\ref{Yketa-phiab-(1)}), 
 one gets that
\begin{align}\label{Yketa-phiab-(1)-ww}
  Y^{(\ww 1)}_{\eta \phi, \gamma \gamma'}(t)
 = y_{\eta,\gamma \gamma' 0} \rho^S_{\gamma \gamma'}(t)
 \equiv  \ww \rho^{S,\eta}_{\gamma \gamma'}(t), 
\end{align} 
 where $\ww \rho^{S,\eta}(t)$ refers to a rescaled RDM in this generic case.
 Following the same procedure as that leading to 
 $\LL_{\alpha \beta}$ in Eq.(\ref{rho(t)-G-t^k-tm-2}), 
 one finds $\ww \LL_{\alpha \beta}(\rho^S)$ written as follows, 
\begin{align} 
  & \ww\LL_{\alpha \beta}(\rho^S) = \sum_{k=1}^{k_{\rm tru}} 
  \sum_{\eta,\gamma,\gamma'}  \tau^{k-1}
 x_{\alpha \beta, \eta}^{(k)  \gamma \gamma'}
 \ww \rho^{S,\eta}_{\gamma \gamma'}.
 \label{ww-Lab-k}
\end{align}
 where $ x_{\alpha \beta, \eta}^{(k)  \gamma \gamma'}$
 were introduced in Sec.\ref{sect-evolv-t0-t} [cf.~Eq.(\ref{Gketa-sum-xYphi})]. 
 This gives rise to a master equation driven by a 
 superoperator $\ww \LL = \sum_{k=1}^{k_{\rm tru}} \ww \LL^{(k)}$.
 Here, $\ww \LL^{(k)}$ of $k=1,2$ have the same expressions
 as given above, meanwhile, those of $k>2$ act as follows, 
\begin{align} 
  & \ww\LL^{(k)}_{\alpha \beta} (\rho^S) = \tau^{k-1} \sum_{\eta} 
  \tr \left( \big[x_{\alpha \beta, \eta}^{(k) }\big]^\dag \ww\rho^{S,\eta}(t)  \right).
 \label{ww-Lab-k-2}
\end{align}
 Here, $\big[x_{\alpha \beta, \eta}^{(k) }\big]$ indicates the matrix of 
 $ x_{\alpha \beta, \eta}^{(k)  \gamma \gamma'}$ with respect to the 
 labels $\gamma$ and $ \gamma'$.

\subsection{An application}\label{sect-refine-application}

 As discussed in Sec.\ref{sect-dephasing}, the basic framework  works well 
 in the study of decoherence under nondissipative interactions.
 However, the basic framework does not work so well for relaxation.
 As an example, let us consider a dissipative interaction for which
 $H^{IS}_{\alpha \alpha} =0$ for all $\alpha$,
 with $k_{\rm tru}=2$ and $d_S =2$. 
 Equation (\ref{LL-k=2}) predicts that, with $\alpha \ne \beta$, 
\begin{align}\label{}
 & \LL_{\alpha \alpha}^{(2)}(\rho^S)
 =  \tau h^{I\E 2}_{ 0} |H^{IS}_{\alpha\beta}|^2 (\rho_{\beta\beta}^S    
 - \rho_{\alpha \alpha}^S).
\label{LL(2)-ab-2}
\end{align}
 A master equation, whose superoperator contains 
 the above $\LL_{\alpha \alpha}^{(2)}$, predicts that 
 a steady state should have the property of 
 $\rho_{\beta\beta}^S = \rho_{\alpha \alpha}^S$.
 In other words, such a master equation may work only for very special 
 steady states, such as a Gibbs state at an infinite temperature.

 Improvement is seen in the refined framework. 
 Now, from Eq.(\ref{LL-k=2-ww}), one finds that
 the superoperator of the master equation should contain
\begin{align}\label{}
 & \ww \LL_{\alpha \alpha}^{(2)}(\rho^S)
 =  \tau  |H^{IS}_{\alpha\beta}|^2 (h^{I\E 2}_{\beta\beta 0}\rho_{\beta\beta}^S    
 - h^{I\E 2}_{\alpha \alpha 0} \rho_{\alpha \alpha}^S).
\label{LL(2)-ab-2-ww}
\end{align}
 This predicts a steady state possessing the following property,
\begin{align}\label{}
 & \frac{\rho_{\alpha \alpha}^S}{\rho_{\beta\beta}^S}
 = \frac{h^{I\E 2}_{\beta\beta 0}}{h^{I\E 2}_{\alpha \alpha 0}},
\label{raa/rbb}
\end{align}
 which may take a value different from $1$.

 One remark: When the value of $\tau$ is not very small, 
 one may need to consider a value of $k_{\rm tru}$ larger than $2$.
 In this case, the superoperator $\ww \LL$ should contain more terms 
 than those discussed above.

\section{About Born and Markov approximations}\label{sect-Born-Mark-appr}

 The ordinary approach to master equation is based
 on the Born and Markov approximations, while, their quantitative 
 justification is still lacking. 
 In this section, within the proposed framework, 
 we justify the Born approximation in a quantitative way
 and argue that time evolution of the RDM should possess a Markovian feature.
 The arguments to be given below are for the averaged RDMs and 
 to stress this point overlines above them will be written explicitly. 

\subsection{Justification of Born approximation}\label{sect-Born-appr}

 For a total system whose initial 
 state is written as $\rho^S(t_0) \otimes \rho^\E_0$,
 where $\rho^\E_0$ represents a thermal state of  the environment,
 the Born approximation assumes that
 the density matrix of the total system,
 indicated as $\rho^{S+\E}(t)$, 
 can be approximated by $\rho^S(t) \otimes \rho^\E_0$.
 Let us still consider a time period $(t_0, t_0+T)$,
 which is divided into a large number $M$ of short intervals
 separated by instatnts $t_m = m\tau $, with a common length $\tau$.

 For the purpose of justifying Born approximation, let us consider
 an arbitrarily chosen time $t_{m_0}$ of $m_0>0$ and $\rho^{S+\E}_{\rm Born}(t)$
 of $t>t_{m_0}$, which is obtained by Schr\"{o}dinger evolution from
 the state of $\rho^{S+\E}_{\rm Born}(t_{m_0}) = \rho^S(t_{m_0}) \otimes \rho^\E_0$.
 Below, we are to show that the true averaged RDM $\ov\rho^S(t)$ of $t>t_{m_0}$
 can be approximated by the averaged RDM $\ov\rho^S_{\rm Bn}(t)$,
 which is computed from $\rho^{S+\E}_{\rm Born}(t)$ with  
 $\rho^S_{\rm Bn}(t) = \tr_\E \rho^{S+\E}_{\rm Born}(t)$.

 We note that Eqs.(\ref{drho-G-t^k-tm}) and (\ref{LL-ab-k}) are still valid
 in the refined framework.
 Then, for $m>m_0$, making use of Eq.(\ref{ww-Lab-k}),  one gets that
\begin{align}
 & \ov\rho^S_{\alpha \beta}(t_{m+1})  - \ov\rho^S_{\alpha \beta}(t_{m})
 \simeq
 \sum_{k=1}^{k_{\rm tru}} 
  \sum_{\eta,\gamma,\gamma'}  \tau^{k}
 x_{\alpha \beta, \eta}^{(k)  \gamma \gamma'}
 \ov{\ww \rho}^{S,\eta}_{\gamma \gamma'} (t_{m}).
 \label{rho(t)-G-t^k-t1-ww}
\end{align}

 To study $\rho^S_{\rm Bn}(t)$, let us write $\rho^{S+\E}_{\rm Born}(t_{m_0})$ 
 in a diagonal form, which is always possible for a Hermitian matrix,
 that is, 
\begin{align}
 & \rho^{S+\E}_{\rm Born}(t_{m_0}) =\sum_p |\Psi_p(t_{m_0})\ra \la \Psi_p(t_{m_0})|,
\end{align}
 whose components are labelled by $p$.
 Expanding each $|\Psi_p(t_{m_0})\ra$ with respect to the system's 
 eigenstates $|\alpha\ra$, one writes
\begin{align}
 |\Psi_p(t_{m_0})\ra = \sum_{\alpha} |\alpha \ra |\Phi^\E_{p\alpha}(t_{1})\ra,
\end{align}
 where $|\Phi^\E_{p\alpha}(t_{1})\ra$ are the corresponding environmental branches. 
 Clearly, the time evolution of $\rho^{S+\E}_{\rm Born}(t)$ is given by 
 Schr\"{o}dinger evolution of $|\Psi_p(t)\ra$ from $|\Psi_p(t_{m_0})\ra$.
 In terms of the environmental branches $|\Phi^\E_{p\alpha}(t)\ra$, 
 the total state is written as 
\begin{align}
 & \rho^{S+\E}_{\rm Born}(t) =\sum_{p,\alpha,\beta} |\alpha \ra \la \beta| 
 \times |\Phi^\E_{p\alpha}(t)\ra  \la \Phi^\E_{p\beta}(t)|.
\end{align}
 It is easy to check that elements of $\rho^S_{\rm Bn}$ have 
 the following expression, 
\begin{equation}\label{rhoBn-abt}
 \rho^S_{{\rm Bn}, \alpha \beta}(t) = \sum_p \rho^S_{{\rm Bn}, p\alpha \beta}(t),
\end{equation}
 where
\begin{equation}\label{rhoab-phiba-tr}
 \rho^S_{{\rm Bn}, p \alpha \beta}(t) = \la \Phi^\E_{p\beta}(t)|\Phi^\E_{p\alpha}(t) \ra.
\end{equation}

 With $\rho^\E_0$ as a thermal state 
 in $\rho^{S+\E}_{\rm Born}(t_{m_0}) = \rho^S(t_{m_0}) \otimes \rho^\E_0$,
 previous discussions for the derivation of master
 equation from the initial state $|\Psi(t_0)\ra$ are still valid,
 if $|\Psi_p(t_{m_0})\ra$ is taken as the initial state for each label $p$.
 This implies validity of Eq.(\ref{rho(t)-G-t^k-t1-ww}) 
 for $\rho^S_{{\rm Bn}, p \alpha \beta}$.
 One key point lies in that, with only contributions of terms of $l=1$ included,
 the rhs of Eq.(\ref{rho(t)-G-t^k-t1-ww}) is a linear function of the RDM.
 As a consequence, taking summation over $p$ on both sides of an equality
 which is similar to Eq.(\ref{rho(t)-G-t^k-t1-ww}) but for 
 $\rho^S_{{\rm Bn}, p \alpha \beta}$, one gets that
\begin{align} \notag
 \ov \rho^S_{{\rm Bn}, \alpha \beta} & (t_{m+1})   \simeq
 \ov\rho^S_{{\rm Bn}, \alpha \beta}(t_{m})
 \\ &  + \sum_{m=0}^{M-1}  \sum_{k=1}^{k_{\rm tru}} 
  \sum_{\eta,\gamma,\gamma'}  \tau^{k}
 x_{\alpha \beta, \eta}^{(k)  \gamma \gamma'}
 \ov{\ww \rho}^{S,\eta}_{{\rm Bn},\gamma \gamma'} (t_{m}).
 \label{rho(t)-G-t^k-t1-ww-Bn}
\end{align}
 Comparing Eq.(\ref{rho(t)-G-t^k-t1-ww}) and Eq.(\ref{rho(t)-G-t^k-t1-ww-Bn})
 and noting that $\rho^S_{{\rm Bn}}(t_{m_0}) = \rho^S(t_{m_0})$, 
 one sees that $\ov\rho^S(t_m)$ of $m>m_0$ 
 indeed can be approximated by $\ov\rho^S_{{\rm Bn}}(t_m)$.
 That is, the Born approximation is applicable in an effective and average sense.

\subsection{Markovian feature of the RDM evolution}\label{sect-Mark-appr}

 In a generic perspective of Markovian process, description of 
 a system at a given time is sufficient for determining its variation with time, 
 not needing information about the past. 
 Within the proposed framework, variation of the averaged RDM 
 in fact shows  a Markovian feature, as discussed below.

 As discussed in Sec.\ref{sect-evolv-t0-t}, change of the exact
 RDM of $S$ from an initial time $t_0$ to a later time $t$
 approximately has a Markovian feature, if the following
 conditions are satisfied: (i) the ETH ansatz for the environment, 
 (ii) approximation of small slope 
 [i.e., sufficient smallness of the slopes $y'_{\eta 0}$ for 
 $k \le k_{\rm tru}(t)$], and (iii) initial thermal state of the environment.
 However, starting from times later than $t_0$, 
 since the time evolution may bring phase correlations among
 the environmental branches, change of the exact RDM is no longer Markovian.

 For times later than $t_0$, it is the averaged RDM whose variation may show
 a Markovian feature, as discussed in Sec.\ref{sect-b-form-ME}.
 In fact, this point is seen clearly in Eq.(\ref{rho(t)-G-t^k-t1-ww}),
 which describes a Markovian process
 with $y_{\eta,\gamma \gamma' 0}$ in $\ww \rho^{S,\eta}_{\gamma \gamma'}(t)$
 taken as constant parameters [cf.~Eq.(\ref{Yketa-phiab-(1)-ww})].
 Roughly speaking, chaotic dynamics of the environment may induce 
 decay of the phase correlations and, as a consequence, 
 time evolution of the RDM may show a fluctuation feature, 
 whose averaged behavior is Markovian. 

\section{Summary and discussions}\label{sect-conclusion}

 In this paper, a framework is proposed for the study of a generic small quantum system $S$,
 which is locally coupled to an environment as
 a huge many-body quantum chaotic system.
 The RDM of $S$ is studied by an approach employing environmental branches, 
 whose overlaps gives elements of the RDM.
 Initially, the environmental branches possess no correlation 
 with the interaction Hamiltonian, 
 e.g., lying in a thermal state.

 The framework is based on three assumptions,
 which are due to the chaotic dynamics of the environment.
 That is, 
 (i) the ETH ansatz; (ii) the approximation of small slope,
 which loosely speaking assumes sufficient smallness of the slopes of the 
 diagonal functions in the ETH ansatz for related operators
 \footnote{The ETH ansatz conjectures smallness of the slopes 
 in a qualitative way.};
and (iii) the approximation of branch-correlation decay, 
 which loosely speaking assumes that certain type of phase correlation among 
 environmental branches decays with the time passing.

 Specifically, the approach consists of three steps as summarized below, 
 concerning the evolution of the system $S$ within a time period $[0,T]$ of interest.
 Firstly, the period $[0,T]$ is divided into a series of short
 time intervals with length $\tau$.
 Within each short interval,
 time evolution of the RDM of $S$ is studied by making use of 
 a formal solution derived for the environmental branches, 
 whose Taylor expansion is effectively truncated at 
 a $k_{\rm tru}$th-order term.
 Secondly, the resulting expression for RDM is simplified
 by making use of the ETH ansatz and the approximation of small slope.
 Thirdly, description for the RDM's evolution in the whole period $[0,T]$ 
 is further simplified by the approximation of branch-correlation decay.
 The above procedures give rise to a master equation.

 As an application of the above discussed framework, 
 the case of $k_{\rm tru}=2$
 has been studied and a master equation has been derived.
 Under a generic nondissipative interaction,
 the master equation predicts decoherence
 under a generic many-body quantum chaotic environment. 
 The predicted decoherence rate reduces to what is known in the special 
 situation in which the RMT (random-matrix theory) is applicable, 
 if an appropriate value of $\tau$ is taken.

 Within the proposed framework, 
 the Born approximation, which is used in the 
 ordinary derivation of master equation without quantitative justification,
 can be justified in a quantitative way.
 And, variation of the RDM is shown to possess a 
 Markovian feature in an average and effective sense.
 Breakdown of the three assumptions discussed above may 
 bring effects that are usually under the name of nonMarkovian.

 Finally, in the master equation gotten in the (refined) framework, 
 the action of the superoperator $\ww\LL$ results in 
 a function of  $ \ww \rho^{S,\eta}_{\gamma \gamma'}(t)
 = y_{\eta,\gamma \gamma' 0} \rho^S_{\gamma \gamma'}(t)$,
 in which the RDM elements are always accompanied by 
 the parameters $y_{\eta,\gamma \gamma' 0}$.
 Loosely speaking, each parameter $y_{\eta,\gamma \gamma' 0}$ represents 
 certain averaged impact of the environmental part of the interaction
 Hamiltonian on the central system.
 Influence of this feature in the evolution of RDM 
 is a topic worth investigation in future.
 \footnote{Whether such a superoperator $\ww \LL$ may be 
 written in a Lindblad form is a mathematical problem, which 
 is expected to be discussed in a different paper.}

\acknowledgements

 This work was partially supported by the Natural Science Foundation of China under Grant
 Nos.~12175222, 92565306, and 11775210.

\appendix

\section{Properties of the operators $Y_\eta$}\label{app-argu-Gk}

 In this appendix, we discuss a method of computing the operators $Y_\eta$
 and discuss their feature with respect to locality.
 To achieve this goal, one needs to write $\rho^S_{\alpha \beta}(t)$
 in the form in Eq.(\ref{drho-G-t^k}), with the G-functions written in the form 
 in Eqs.(\ref{Gk-sum-xYphi})-(\ref{Gketa-sum-xYphi}).
 Making use of
 Eqs.(\ref{rho-tr-PhiE-ab}) and (\ref{|Phi>-exp})-(\ref{<Phi|-exp}), one gets that
\begin{align}\label{rho-exp-phi-app}
 & \rho^S_{\alpha \beta}(t) =
  \big[ \la \phi^\E(t_0)| \big] e^{i \M (t-t_0)}  [P]  e^{-i \M (t-t_0)}\big[ |\phi^\E(t_0)\ra \big],
\end{align}
 where $[P]$ is a $\alpha$-matrix whose elements are determined by
 $P_{\gamma \gamma'} \equiv \delta_{\gamma \beta } \delta_{\gamma' \alpha}$.

 To compute $Y_\eta$ from Eq.(\ref{rho-exp-phi-app}),
 one may make use of the following mathematical formula for two matrices $A$ and $B$,
\begin{align}\label{eABe-A}
 e^A B e^{-A} = \sum_{n=0}^\infty \frac{1}{n!} C_n,
\end{align}
 where $C_n$ is determined by the following iteration relation,
\begin{align}\label{}
 C_n = [A, C_{n-1}], \quad \text{with $C_0 = B$.}
\end{align}
 Indeed, taking $A = i \M (t-t_0)$ and $B = [P]$, where $\M$ is defined in Eq.(\ref{M-matrix}), namely,
\begin{align}\notag
 & \M = [H^S] +  H^\E [I] +  H^{I\E}[H^{IS}],
\end{align}
 one sees that 
\begin{align}\notag
 e^{i \M (t-t_0)}  [P]  e^{-i \M (t-t_0)} = \sum_{n=0}^\infty \frac{1}{n!} C_n.
\end{align}
 It is easy to see that $C_n$ has the following dependence on $t$,
\begin{align}\label{}
 C_n = i^n (t-t_0)^n  \ww C_{n},
\end{align}
 where $\ww C_n$ is determined by the iteration relation 
 of $\ww C_n = [\M, \ww C_{n-1}]$ with $\ww C_0 = [P]$.
 Thus, $n$ is equal to the label $k$.

 Let us first discuss the case of $k=1$.
 It is straightforward to get that
\begin{align}\label{}\notag
  \ww C_1 = [\M, [P]] & = [[H^S], [P]] +  H^{I\E}[[H^{IS}], [P]]
 \\ & \equiv [d_{1}] + [d_{2}] H^{I\E}. \label{wwC1}
\end{align}
 Here, we have introduced parameter matrices $[d_{q}]$ of $q=1,2$ (as $\alpha$-matrices)
 for the purpose of writing the result in a concise way,
 which can be computed directly from the third equality in Eq.(\ref{wwC1}).
 Thus, from $C_1$, one finds that $Y_1 = I$ and $Y_2 = H^{I\E}$ for $k=1$.

 Then, we discuss $k=2$.
 It is easy to find that $\ww C_2 = [\M, \ww C_1]$ has the following form,
\begin{align}\label{}
 & \ww C_2  = [d_{3}] + [d_{4}] H^{I\E} + [d_{5}] H^{I\E,\E} + [d_{6}] (H^{I\E})^2.
\end{align}
 Thus, for $k=2$, beside $Y_1 = I$ and $Y_2 = H^{I\E}$, one finds that
 $Y_3 = H^{I\E,\E} \equiv [H^{I\E}, H^\E]$ and $Y_4 = (H^{I\E})^2$.
 Thus, Eq.(\ref{Yk-01}) is proved.

 Proceeding with the above procedure, one may find expressions of $Y_\eta$ for $k>2$.
 In particular, regarding locality, one finds the following property.
 That is, among the operators $Y_\eta$ that appear at a given value of $k$,
 the one possessing the widest interaction scope is
 the operator that is constructed from commutators built from one $H^{I\E}$ and a number
 $(k-1)$ of $H^\E$, labelled by ${\eta_{\rm M}}$, namely,
\begin{align}\label{}
 & Y_{\eta_{\rm M}} = [ \cdots [ [H^{I\E}, H^\E], H^\E] \cdots , H^\E].
\end{align}
 Then, for $k$ not large, one sees that
 the operators $Y_\eta$ are either the identity operator $I$ or local operators.
 Clearly, within a fixed and finite time period, in the thermodynamic limit of $N\to \infty$, 
 all the operators $Y_\eta$ which are effectively relevant are either local operators or $I$.
 In this limit, one may take $k_{\rm ETH}$ as large as one would like, without any upper bound.

\section{Computation of $G^{(k)}_{\alpha \beta}(t_0)$ of $k=1,2$}\label{app-derive-Gk}

 In this appendix, we derive expressions of $G^{(k)}_{\alpha \beta}(t_m)$,
 as the expansion coefficients of $\rho^S_{\alpha \beta}(t_{m+1})  
 - \rho^S_{\alpha \beta}(t_{m})$
 in the power of $(t_{m+1}-t_m)^{k}$ [cf.~Eq.(\ref{drho-G-t^k})].
 More exactly, we derive expressions given in Eqs.(\ref{G1-0})-(\ref{G1-0-4}) for $k=1$
 and in Eqs.(\ref {G21-Q-2HIS})-(\ref{G21-Q-2HIS-4}) for $k=2$ in the main text.
 For brevity, we set $t_m=0$ and write $t_{m+1}$ as $t$; thus, the expansion is in the power of $t^k$.
 Also for brevity, repeated index imply summation over the index 
 in this section, unless otherwise stated.

 Writing the expansion of the component $\big[|\phi^\E(t)\ra \big]_\alpha $
 [see Eq.(\ref{|Phi>-exp})] explicitly, one has
\begin{gather} \notag
 \big[|\phi^\E(t)\ra \big]_\alpha =
 \sum_n \frac{(-i t)^n}{n!} (\M^n \big[ |\phi^\E(0)\ra \big] )_\alpha
 \\ \notag =  |\phi_\alpha^\E(0)\ra  -i t \M_{\alpha \gamma}  |\phi_\gamma^\E(0)\ra
   - \frac 12 t^2 \M^2_{\alpha \gamma}  |\phi_\gamma^\E(0)\ra
 \\ +i \frac 16 t^3 \M^3_{\alpha \gamma}  |\phi_\gamma^\E(0)\ra
 + \frac 1{24} t^4 \M^4_{\alpha \gamma}  |\phi_\gamma^\E(0)\ra  + \cdots;  \label{|Phi>-exp-app}
\end{gather}
 and, similarly  for the bra,
\begin{gather} \notag \big[\la \phi^\E(t)| \big]_\beta =
 \sum_n \frac{(i t)^n}{n!} (\big[ \la \phi^\E(0)| \big] \M^n )_\beta
 \\  \notag = \la \phi_\beta^\E(0)|  + i t   \la \phi_\gamma^\E(0)|  \M_{ \gamma\beta}
   - \frac 12 t^2  \la \phi_\gamma^\E(0)|  \M^2_{ \gamma \beta}
 \\ -i \frac 16 t^3 \la \phi_\gamma^\E(0)|  \M^3_{ \gamma \beta}
 + \frac 1{24} t^4 \la \phi_\gamma^\E(0)|  \M^4_{\gamma \beta } + \cdots.  \label{<Phi|-exp-app}
\end{gather}
 Substituting the above expansions into Eq.(\ref{rho-tr-PhiE-ab})
 and making use of Eq.(\ref{drho-G-t^k}),
 it is direct to get expressions of $G^{(k)}_{\alpha \beta}(0)$.
 Note that, according to Eq.(\ref{M-matrix}), $\M = [H^S] +  H^\E [I] +  H^{I\E}[H^{IS}]$ and
\begin{align}\label{}
  \M_{\alpha \gamma} = (e^S_{\alpha} +  H^\E) \delta_{\alpha \gamma}  +  H^{I\E} H^{IS}_{\alpha \gamma}.
\end{align}

\begin{widetext}

 For $k=1$, one gets the following term before $t$,
\begin{align}\label{}\notag
  & G^{(1)}_{\alpha \beta}(0) = \la \phi_\beta^\E(0)|  (-i ) \M_{\alpha \gamma} |\phi_\gamma^\E(0)\ra
  +i  \la \phi_\gamma^\E(0)|  \M_{ \gamma\beta}  |\phi_\alpha^\E(0)\ra
  \\ & = i (e^S_{\beta} - e^S_{\alpha}) \la \phi_\beta^\E (0) |\phi_\alpha^\E (0) \ra
  + i  \Big( H^{IS}_{\gamma  \beta }
    \bra{\phi_\gamma^\E (0)}H^{I\E} \ket{\phi_\alpha^\E (0) }
   - H^{IS}_{\alpha \gamma  } \bra{\phi_\beta^\E (0)} H^{I\E} \ket{\phi_\gamma^\E (0)} \Big).
\label{G1-0-app}
\end{align}
 This gives Eqs.(\ref{G1-0})-(\ref{G1-0-4}).
 For $k=2$, one gets the following term  before $t^2$,
\begin{align}\label{}
  G^{(2)}_{\alpha \beta}(0) & =
  \la \phi_\gamma^\E(0)|  \M_{ \gamma\beta}  \M_{\alpha \gamma'}  |\phi_{\gamma'}^\E(0)\ra
  - \frac 12 \la \phi_\gamma^\E(0)| \M^2_{ \gamma \beta} |\phi_\alpha^\E(0)\ra
  - \frac 12 \la \phi_\beta^\E(0)|  \M^2_{\alpha \gamma} |\phi_\gamma^\E(0)\ra .
\end{align}

 Below, we derive an explicit expression for $G^{(2)}_{\alpha \beta}(0)$.
 Making use of the following relation,
\begin{gather}
  \M_{ \gamma\beta}  \M_{\alpha \gamma'}
 = \Big(\delta_{\gamma\beta} (e^S_\beta + H^\E) + H^{IS}_{\gamma\beta} H^{I\E} \Big)
  \Big(\delta_{\alpha \gamma'} (e^S_\alpha + H^\E ) + H^{IS}_{\alpha \gamma'} H^{I\E} \Big),
\end{gather}
 one gets
\begin{gather*}
    \la \phi_\gamma^\E(0)|  \M_{ \gamma\beta}  \M_{\alpha \gamma'}  |\phi_{\gamma'}^\E(0)\ra
 =   \la \phi_\gamma^\E(0)|
 \Big( \delta_{\gamma\beta}  (e^S_\beta + H^\E) \delta_{\alpha \gamma'}  (e^S_\alpha + H^\E)
  + \delta_{\gamma\beta}  (e^S_\beta + H^\E) H^{IS}_{\alpha \gamma'} H^{I\E}
 \\ + H^{IS}_{\gamma\beta} H^{I\E} \delta_{\alpha \gamma'}  (e^S_\alpha + H^\E)
 + H^{IS}_{\gamma\beta}  H^{I\E} H^{IS}_{\alpha \gamma'} H^{I\E}
 \Big)  |\phi_{\gamma'}^\E(0)\ra
 \\ =  \la \phi^\E_\beta(0)| (e^S_\beta + H^\E)(e^S_\alpha + H^\E) |\phi^\E_{\alpha}(0)\ra
 + H^{IS}_{\alpha \gamma'}  \la \phi^\E_\beta(0)|  (e^S_\beta + H^\E)  H^{I\E}  |\phi^\E_{\gamma'}(0)\ra
 \\ +  H^{IS}_{\gamma\beta} \la \phi^\E_\gamma(0)| H^{I\E}  (e^S_\alpha + H^\E)  |\phi^\E_{\alpha}(0)\ra
 + H^{IS}_{\gamma\beta}  H^{IS}_{\alpha \gamma'} \la \phi^\E_\gamma(0)|  H^{I\E} H^{I\E} |\phi^\E_{\gamma'}(0)\ra
 \\ =  \la \phi^\E_\beta(0)| (H^\E)(H^\E) |\phi^\E_{\alpha}(0)\ra
 + H^{IS}_{\alpha \gamma'}  \la \phi^\E_\beta(0)|  (H^\E)  H^{I\E}  |\phi^\E_{\gamma'}(0)\ra
 \\ +  H^{IS}_{\gamma\beta} \la \phi^\E_\gamma(0)| H^{I\E}  (H^\E)  |\phi^\E_{\alpha}(0)\ra
 + H^{IS}_{\gamma\beta}  H^{IS}_{\alpha \gamma'} \la \phi^\E_\gamma(0)|  H^{I\E} H^{I\E} |\phi^\E_{\gamma'}(0)\ra
 \\ + \la \phi^\E_\beta(0)| (e^S_\beta e^S_\alpha + e^S_\beta H^\E + e^S_\alpha H^\E) |\phi^\E_{\alpha}(0)\ra
 + H^{IS}_{\alpha \gamma'}  \la \phi^\E_\beta(0)|  e^S_\beta  H^{I\E}  |\phi^\E_{\gamma'}(0)\ra
 +  H^{IS}_{\gamma\beta} \la \phi^\E_\gamma(0)| H^{I\E}  e^S_\alpha |\phi^\E_{\alpha}(0)\ra.
\end{gather*}

 Furthermore, we note that
\begin{gather}\notag
 \M^2_{ \gamma \beta} = \M_{ \gamma \eta}\M_{ \eta \beta}
 = (\delta_{\gamma \eta}  (e^S_\gamma + H^\E) + H^{IS}_{\gamma \eta} H^{I\E})
 (\delta_{\eta \beta}  (e^S_\beta + H^\E) + H^{IS}_{\eta \beta} H^{I\E})
 \\ \notag = \delta_{\gamma \beta} (e^S_\beta + H^\E)(e^S_\beta + H^\E)
 + H^{IS}_{\gamma \beta} (e^S_\gamma + H^\E)  H^{I\E}
 + H^{IS}_{\gamma \beta} H^{I\E} (e^S_\beta + H^\E) + H^{IS}_{\gamma \eta} H^{IS}_{\eta \beta} (H^{I\E})^2
 \\ \notag = \delta_{\gamma \beta} (H^\E)^2  + H^{IS}_{\gamma \beta} \{ H^\E,  H^{I\E} \}
 + (H^{IS})^2_{\gamma \beta} (H^{I\E})^2
 \\ + \delta_{\gamma \beta} (e^S_\beta e^S_\beta + 2e^S_\beta H^\E)
 + H^{IS}_{\gamma \beta} e^S_\gamma H^{I\E} + H^{IS}_{\gamma \beta} H^{I\E} e^S_\beta
 \quad \text{(no summation over $\beta$ and $\gamma$).}
\end{gather}
Thus, one gets that
\begin{gather*}
   \la \phi_\gamma^\E(0)|  \M^2_{ \gamma \beta}  |\phi_\alpha^\E(0)\ra
 =   \la \phi_\gamma^\E(0)| \Big(
  \delta_{\gamma \beta} (H^\E)^2  + H^{IS}_{\gamma \beta} \{ H^\E,  H^{I\E} \} + (H^{IS})^2_{\gamma \beta} (H^{I\E})^2
 \\ + \delta_{\gamma \beta} (e^S_\beta e^S_\beta + 2e^S_\beta H^\E)
 + H^{IS}_{\gamma \beta} e^S_\gamma H^{I\E} + H^{IS}_{\gamma \beta} H^{I\E} e^S_\beta
 \Big) |\phi_\alpha^\E(0)\ra
 \\ =   ( \la \phi^\E_\beta(0)| (H^\E)^2  |\phi^\E_\alpha(0)\ra
 + \la \phi^\E_\gamma(0)| H^{IS}_{\gamma \beta} \{ H^\E,  H^{I\E} \}  |\phi^\E_\alpha(0)\ra
 + \la \phi^\E_\gamma(0)| (H^{IS})^2_{\gamma \beta} (H^{I\E})^2  |\phi^\E_\alpha(0)\ra  )
 \\ +    \la \phi^\E_\beta(0)| (e^S_\beta e^S_\beta + 2e^S_\beta H^\E) |\phi^\E_\alpha(0)\ra
 + \la \phi^\E_\gamma(0)| H^{IS}_{\gamma \beta} (e^S_\gamma + e^S_\beta ) H^{I\E}   |\phi^\E_\alpha(0)\ra.
\end{gather*}

 To get $\la \phi_\beta^\E(0)|  \M^2_{ \alpha \gamma}  |\phi_\gamma^\E(0)\ra $, one may make
 use of the complex conjugate of the above quantity, which is obtained by reversing all ket-bras from left to right,
 then, exchanging $\alpha$ and $\beta$ ($\alpha \leftrightarrow \beta$), i.e.,
\begin{align}\label{}
   \la \phi_\gamma^\E(0)|  \M^2_{ \gamma \beta}  |\phi_\alpha^\E(0)\ra
   \to    \la \phi_\alpha^\E(0)|  \M^2_{ \beta \gamma}  |\phi_\gamma^\E(0)\ra
   \to    \la \phi_\beta^\E(0)|  \M^2_{ \alpha \gamma}  |\phi_\gamma^\E(0)\ra.
\end{align}
 The first step gives
\begin{align}\label{}\notag
 &   ( \la \phi_\alpha^\E(0)| (H^\E)^2  |\phi^\E_\beta(0)\ra
 + \la \phi_\alpha^\E(0)| H^{IS}_{\beta \gamma } \{ H^\E,  H^{I\E} \}  |\phi^\E_\gamma(0)\ra
 + \la \phi_\alpha^\E(0)| (H^{IS})^2_{\beta \gamma } (H^{I\E})^2  |\phi^\E_\gamma(0)\ra  )
 \\  & +   ( \la \phi_\alpha^\E(0)| (e^S_\beta e^S_\beta + 2e^S_\beta H^\E) |\phi^\E_\beta(0)\ra
 + \la \phi_\alpha^\E(0)| H^{IS}_{\beta \gamma } (e^S_\gamma + e^S_\beta ) H^{I\E}   |\phi^\E_\gamma(0)\ra.
\end{align}
 And, the second step gives
\begin{align}\label{}\notag
 &   ( \la \phi_\beta^\E(0)| (H^\E)^2  |\phi_\alpha^\E(0)\ra
 + \la \phi_\beta^\E(0)| H^{IS}_{\alpha \gamma } \{ H^\E,  H^{I\E} \}  |\phi^\E_\gamma(0)\ra
 + \la \phi_\beta^\E(0)| (H^{IS})^2_{\alpha \gamma } (H^{I\E})^2  |\phi^\E_\gamma(0)\ra  )
 \\  & +   \la \phi_\beta^\E(0)| (e^S_\alpha e^S_\alpha + 2e^S_\alpha H^\E) |\phi^\E_\alpha(0)\ra
 + \la \phi_\beta^\E(0)| H^{IS}_{\alpha \gamma } (e^S_\gamma + e^S_\alpha ) H^{I\E}   |\phi^\E_\gamma(0)\ra.
\end{align}

 We divide $G^{(2)}_{\alpha \beta}(0)$ into two parts, $G^{(2)}_{\alpha \beta}(0) = g_1 + g_2$,
 where $g_1$ contains the system $S$'s energy and $g_2$ does not.
 One finds that
\begin{gather*}
 g_1 =  \la \phi^\E_\beta(0)| (e^S_\beta e^S_\alpha + e^S_\beta H^\E
 + e^S_\alpha H^\E) |\phi^\E_{\alpha}(0)\ra
 + H^{IS}_{\alpha \gamma}  \la \phi^\E_\beta(0)|  e^S_\beta  H^{I\E}  |\phi^\E_{\gamma}(0)\ra
 +  H^{IS}_{\gamma\beta} \la \phi^\E_\gamma(0)| H^{I\E}  e^S_\alpha |\phi^\E_{\alpha}(0)\ra
 \\ - \frac 12 \Big(    \la \phi^\E_\beta(0)| (e^S_\beta e^S_\beta + 2e^S_\beta H^\E) |\phi^\E_\alpha(0)\ra
 + \la \phi^\E_\gamma(0)| H^{IS}_{\gamma \beta} (e^S_\gamma + e^S_\beta ) H^{I\E}   |\phi^\E_\alpha(0)\ra \Big)
 \\     - \frac 12 \Big( \la \phi_\beta^\E(0)| (e^S_\alpha e^S_\alpha + 2e^S_\alpha H^\E) |\phi^\E_\alpha(0)\ra
 + \la \phi_\beta^\E(0)| H^{IS}_{\alpha \gamma } (e^S_\gamma + e^S_\alpha ) H^{I\E}   |\phi^\E_\gamma(0)\ra \Big)
 \\ =  \la \phi^\E_\beta(0)| (e^S_\beta e^S_\alpha - \frac 12 e^S_\beta e^S_\beta -\frac 12 e^S_\alpha e^S_\alpha )
 |\phi^\E_{\alpha}(0)\ra
 + H^{IS}_{\alpha \gamma}  \la \phi^\E_\beta(0)| ( e^S_\beta -\frac 12
   (e^S_\gamma + e^S_\alpha ) ) H^{I\E}  |\phi^\E_{\gamma}(0)\ra
 \\ +  H^{IS}_{\gamma\beta} \la \phi^\E_\gamma(0)| H^{I\E}  (e^S_\alpha -\frac 12
 (e^S_\gamma + e^S_\beta ) )|\phi^\E_{\alpha}(0)\ra.
\end{gather*}
 This gives
\begin{gather} 
 g_1 =
 - \frac 12  (e^S_\beta - e^S_\alpha)^2  \rho^S_{\alpha \beta}(0)
 +  H^{IS}_{\alpha \gamma} (e^S_\beta -\frac 12   (e^S_\gamma + e^S_\alpha ) ) H^{I\E}_{\phi, \gamma \beta}(0)
  + H^{IS}_{\gamma\beta}(e^S_\alpha -\frac 12  (e^S_\gamma + e^S_\beta ) ) H^{I\E}_{\phi, \alpha\gamma}(0).
   \label{app-G2-l=0}
\end{gather}
 Note that 
\begin{align}\label{}
  (e^S_\beta - e^S_\alpha)^2  \rho^S_{\alpha \beta}(0)\ra
 =   \la \beta| H^S [H^S,\rho^S] |\alpha\ra,
\end{align}
 where the following relation has been used
\begin{gather}\notag
  \la \beta| H^S [H^S,\rho^S] |\alpha\ra
   \notag =  \la \beta| (H^S (H^S \rho^S - \rho^S H^S) - (H^S \rho^S - \rho^S H^S) H^S) |\alpha\ra
 \\  = e_\beta^S e_\beta^S \rho_{\alpha \beta}^S - 2e_\beta^S \rho_{\alpha \beta}^S e_\alpha^S
   +\rho_{\alpha \beta}^S e_\alpha^S e_\alpha^S .
\end{gather}
 For $g_2$, one finds that
\begin{gather*}
 g_2=   \la \phi^\E_\beta(0)| ( H^\E)^2 |\phi^\E_{\alpha}(0)\ra
 + H^{IS}_{\alpha \gamma}  \la \phi^\E_\beta(0)|  H^\E  H^{I\E}  |\phi^\E_{\gamma}(0)\ra
 \\ +  H^{IS}_{\gamma\beta} \la \phi^\E_\gamma(0)| H^{I\E}  H^\E  |\phi^\E_{\alpha}(0)\ra
 + H^{IS}_{\gamma\beta}  H^{IS}_{\alpha \gamma'} \la \phi^\E_\gamma(0)|  H^{I\E} H^{I\E} |\phi^\E_{\gamma'}(0)\ra
 \\ - \frac 12 ( \la \phi^\E_\beta(0)| (H^\E)^2  |\phi^\E_\alpha(0)\ra
 + \la \phi^\E_\gamma(0)| H^{IS}_{\gamma \beta} \{ H^\E,  H^{I\E} \}  |\phi^\E_\alpha(0)\ra
 + \la \phi^\E_\gamma(0)| (H^{IS})^2_{\gamma \beta} (H^{I\E})^2  |\phi^\E_\alpha(0)\ra  )
 \\     - \frac 12 ( \la \phi^\E_\beta(0)| (H^\E)^2  |\phi^\E_\alpha(0)\ra
 + \la \phi^\E_\beta(0)| H^{IS}_{\alpha \gamma} \{ H^\E,  H^{I\E} \} |\phi^\E_\gamma(0)\ra
 + \la \phi^\E_\beta(0)| (H^{IS})^2_{\alpha \gamma} (H^{I\E})^2 |\phi^\E_\gamma(0)\ra ).
\end{gather*}
 This gives
\begin{gather}\notag
  g_2= \frac 12 H^{IS}_{\gamma \beta} \la \phi^\E_\gamma(0)| [H^{I\E}, H^\E ]  |\phi^\E_\alpha(0)\ra
 + \frac 12 \la \phi^\E_\beta(0)| H^{IS}_{\alpha \gamma} [ H^\E,  H^{I\E} ] |\phi^\E_\gamma(0)\ra
 \\ \notag
 + H^{IS}_{\gamma\beta}  H^{IS}_{\alpha \gamma'} \la \phi^\E_\gamma(0)|  (H^{I\E})^2  |\phi^\E_{\gamma'}(0)\ra
 \\ -\frac 12 ( \la \phi^\E_\gamma(0)| (H^{IS})^2_{\gamma \beta} (H^{I\E})^2  |\phi^\E_\alpha(0)\ra
 -\frac 12 \la \phi^\E_\beta(0)| (H^{IS})^2_{\alpha \gamma} (H^{I\E})^2 |\phi^\E_\gamma(0)\ra ).
 \label{G2l-1,4}
\end{gather}

 It is not difficult to see that 
\begin{align}\label{}
 g_1 = G^{(2)}_{\alpha \beta 1} + G^{(2)}_{\alpha \beta 2}, \quad
 g_2 = G^{(2)}_{\alpha \beta 3} + G^{(2)}_{\alpha \beta 4}.
\end{align}
 Then, one gets Eqs.(\ref {G21-Q-2HIS})-(\ref{G21-Q-2HIS-4}).

\end{widetext}


\begin{thebibliography}{99}


\bibitem{breuer2002} H.-P. Breuer and F. Petruccione, \textit{The Theory of Open Quantum Systems}
(Oxford University Press, 2002).


\bibitem{GongJB12} C. K. Lee, J.-S. Cao, J.-B. Gong, Phys. Rev. E {\bf 86}, 021109 (2012).
\bibitem{addis2014coherence} C. Addis, G. Brebner, P. Haikka, and S. Maniscalco, Phys. Rev. A {\bf 89}, 024101 (2014).
\bibitem{roszak2015decoherence} K. Roszak, R. Filip, and T. Novotn\`{y}, Scientific reports {\bf 5}, 9796 (2015).
\bibitem{zhang2015role} Y.-J. Zhang, W. Han, Y.-J. Xia, Y.-M. Yu, and H. Fan, Scientific reports {\bf 5}, 13359 (2015).
\bibitem{cakmak2017} {\ifmmode \mbox{\c{C}}\else \c{C}\fi{}akmak, B. and Manatuly, A.
	and M\"ustecapl\ifmmode \imath \else \i \fi{}o\ifmmode \breve{g}\else \u{g}\fi{}lu, \"O. E.},
Phys. Rev. A {\bf 96}, 032117 (2017)
\bibitem{guarnieri2018steady} G. Guarnieri, M. Kolar, and R. Filip, Phys. Rev. Lett. {\bf 121}, 070401 (2018).


\bibitem{RBC06} D.~Rossini, G.~Benenti, and G.~Casati, Phys.~Rev.~E {\bf 74}, 036209 (2006).
\bibitem{Rigol-AiP16} L. D'Alessio, Y. Kafri, A. Polkovnikov, and M. Rigol, Advances in Physics {\bf 65}, 239 (2016).
\bibitem{Deutch-RPP18} J.M. Deutsch, Rep. Prog. Phys. {\bf 81}, 082001 (2018) (arXiv:1805.01616).

\bibitem{Deutch91} J.M. Deutsch, Phys. Rev. A {\bf 43}, 2046 (1991).
\bibitem{srednicki1994ET} M. Srednicki, Phys. Rev. E {\bf 50}, 888 (1994).
\bibitem{srednicki-JPA96} M. Srednicki, J. Phys. A {\bf 29}, L75 (1996).
\bibitem{srednicki1999ETH} M. Srednicki, J. Phy. A, {\bf 32}, 1163 (1999).

 \bibitem{DonvSMGM-PRB25} P. O’Donovan, P. Strasberg, K. Modi, J. Goold, and M. T. Mitchison,
 Quantum master equation from the eigenstate thermalization hypothesis,
 Phys. Rev. B {\bf 112}, 014312 (2025).

\bibitem{GPSS04} T.~Gorin, T. Prosen, T.H. Seligman, and W.T. Strunz, Phys. Rev. A {\bf 70}, 042105 (2004).
\bibitem{pra08-PS} W.-g.~Wang, J.~B.~ Gong, G.~Casati, and B.~ Li, Phys.~Rev.~A {\bf 77},
 012108 (2008).


 \bibitem{peres84} A.Peres, Phys.Rev. A {\bf 30}, 1610 (1984).
\bibitem{jal2001} R. A. Jalabert and H. M. Pastawski, Phys. Rev. Lett. \textbf{86}, 2490 (2001).
\bibitem{jac2001} Ph. Jacquod, P. G. Silvestrov, and C. W. J. Beenakker, Phys. Rev. E \textbf{64},
 055203(R) (2001).
\bibitem{CT02} N. R. Cerruti and S. Tomsovic, Phys. Rev. Lett. \textbf{88}, 054103 (2002);
J. Phys. A {\bf 36}, 3451 (2003).
 \bibitem{cuc2002} F. M. Cucchietti, C. H. Lewenkopf, E. R. Mucciolo, H. M. Pastawski, and
 R. O. Vallejos, Phys. Rev. E \textbf{65}, 046209 (2002).
 \bibitem{prosen2002} T. Prosen and M. \v{Z}nidari\v{c}, J. Phys. A \textbf{35}, 1455 (2002).
 \bibitem{STB03} P. G.~Silvestrov, J.~Tworzyd{\l}o, and C. W. J.~Beenakker,
   Phys.~Rev.~E {\bf 67}, 025204(R) (2003).
\bibitem{WCL04} W.-g. Wang, G. Casati, and B. Li, Phys. Rev. E {\bf 69} , 025201(R) (2004).
\bibitem{WL05} W.-g. Wang and B. Li Phys. Rev. E \textbf{71}, 066203 (2005).
 \bibitem{GPSZ06} T.~Gorin, T.~Prosen, T.H.~Seligman, and M.~\v{Z}nidari\v{c},
 Phys.~Rep.~{\bf 435}, 33 (2006).



\bibitem{Wilkinson} M. Wilkinson, J. Phys. A {\bf 21}, 1173 (1988).
\bibitem{Eckhardt95} B. Eckhardt, S. Fishman, J. Keating, O. Agam, J. Main, and K.
M\"{u}ller, Phys. Rev. E {\bf 52}, 5893 (1995).
\bibitem{Eckhardt-Main95} B. Eckhardt and J. Main, Phys. Rev. Lett. {\bf 75}, 2300 (1995).
\bibitem{Castella_1996} H. Castella and X. Zotos, Phys. Rev. B {\bf 54}, 4375 (1996).
\bibitem{Sred-H-98} S. Hortikar and M. Srednicki,
 Phys. Rev. E {\bf 57}, 7313 (1998).

\bibitem{pra22-PB-ETH} H. Yan, J. Wang, and W.-g. Wang, Phys. Rev. A {\bf 106}, 042219 (2022).

\bibitem{Steinigeweg_2013} R. Steinigeweg, J. Herbrych, and P. Prelovsek, Phys. Rev. E {\bf 87}, 012118 (2013).
\bibitem{Rigol_2017} R. Mondaini and M. Rigol, Phys. Rev. E {\bf 96}, 012157 (2017).
\bibitem{Dymarsky_Gemmer_2020} J. Richter, A. Dymarsky, R. Steinigeweg,
and J. Gemmer, Phys. Rev. E {\bf 102}, 042127 (2020).



\bibitem{Jiaozi22ETHdevi} J. Wang, M. H. Lamann,  J. Richter, R. Steinigeweg,
 A.  Dymarsky,  and J. Gemmer, \prl {\bf 128}, 180601 (2022).
 \bibitem{Dymarsky22BoundETH} A.  Dymarsky  \prl {\bf 128}, 190601 (2022).
 \bibitem{Vidmar20-prl} M. Mierzejewski  and L. Vidmar, \prl {\bf 124}, 040603 (2020).
 \bibitem{Nussinov22} Z. Nussinov and S. Chakrabarty, Ann. Phys. {\bf 443} 168970 (2022).
 \bibitem{Jiaozi-PRX25} L. Capizzi, J. Wang, X. Xu, L. Mazza, and D. Poletti, 
  Phys. Rev. X, {\bf 15}, 011059 (2025). arXiv:2405.16975 (2024).
\bibitem{pre25-ETHsc} X. Wang and W.-g. Wang, Phys. Rev. E {\bf 112}, 054215 (2025).
arXiv:2210.13183.
\bibitem{ctp25-obs-corr} X. Wang, J. Wang, and W.-g. Wang,
Comm. Theo. Phys. {\bf 77}, 125601 (2025).
\bibitem{WW-ETH-conf} X. Wang and W.-g. Wang,
 arXiv: 2509.24490.



 \bibitem{KIH-pre14} H. Kim, T. N. Ikeda, and D. A. Huse, {\pre} {\bf 90}, 052105 (2014).
 \bibitem{Rigol-prb20}  M. Brenes, J. Goold, and M. Rigol, {\prb} {\bf 102}, 075127 (2020).




\end{thebibliography}
\end{document}